\documentclass[aps,amsmath,amssymb,nofootinbib,notitlepage,superscriptaddress,twocolumn]{revtex4-1}
\pdfoutput=1

\usepackage{aas_macros,mathtools,esint,graphicx,overpic,xcolor,multirow,tabularx}
\usepackage[bottom]{footmisc}
\usepackage[colorlinks=true]{hyperref}
\definecolor{red}{RGB}{228,26,28}
\definecolor{green}{RGB}{77,175,74}
\definecolor{blue}{RGB}{55,126,184}
\definecolor{purple}{RGB}{152,78,163}

\let\originalleft\left
\let\originalright\right
\renewcommand{\left}{\mathopen{}\mathclose\bgroup\originalleft}
\renewcommand{\right}{\aftergroup\egroup\originalright}

\begin{document}
\title{Net Charge Accretion in Magnetized Kerr Black Holes}
\author{Ethan~Berreby}
\email{ethanberreby@campus.technion.ac.il}
\affiliation{Department of Physics, Technion, Haifa 32000, Israel}

\author{Avner~Okun}
\email{avnerok@campus.technion.ac.il}
\affiliation{Department of Physics, Technion, Haifa 32000, Israel}

\author{Shahar~Hadar}
\email{shaharhadar@sci.haifa.ac.il}
\affiliation{
Department of Mathematics and Physics, University of Haifa at Oranim, Kiryat Tivon 3600600, Israel}
\affiliation{Haifa Research Center for Theoretical Physics and Astrophysics, University of Haifa, Haifa 3498838, Israel}

\author{Amos~Ori}
\email{amos@physics.technion.ac.il}
\affiliation{Department of Physics, Technion, Haifa 32000, Israel}

\begin{abstract}
We investigate the charging process of a rotating Kerr black hole
of mass $M$ and angular momentum $J$ immersed in a stationary, axisymmetric,
asymptotically uniform magnetic field of strength $B_{0}$. In Wald's
classic analysis \cite{Wald1974}, which was based on the assumption of vanishing injection energy, the black hole was predicted to acquire a universal
``saturation charge" $Q_{\mathrm{w}}=2B_{0}J$. However, the physical mechanism that sets the saturation
charge must ultimately be governed by the competition between the
absorption rates of positively and negatively charged particles.
Motivated by this observation, we revisit the problem in the framework
of a simple accretion model, where two dilute, equivalent fluxes of
charged particles of opposite signs are injected from infinity along
the magnetic field lines. The problem then reduces to that of individual
particle motion in the electromagnetic field of the magnetized Kerr black hole.
Using a combination of 
numerical and analytical tools, 
we determine the domains of absorption and establish both lower and upper bounds on the corresponding absorption
cross sections. 
At $Q=Q_\mathrm{w}$ these bounds reveal a systematic
difference between the two charge signs. In particular, for sufficiently
strong magnetic fields, the lower bound on the absorption cross section for the
``attracted" charge exceeds the
upper bound for the ``repelled" one. This charge accretion imbalance (which we find to become extreme at the limit
of large $B_{0}$) indicates a persistent net charge accretion at
$Q=Q_{\mathrm{w}}$, implying that the actual saturation charge must
differ from Wald's charge $Q_{\mathrm{w}}$.
\end{abstract}

\maketitle

\section{Introduction}

\label{chap:intro}

When a rotating black hole (BH) 
is immersed in an environmental magnetic
field, the strong curvature distorts the electromagnetic
field, creating an electric field -- which may be strong in
the close vicinity of the BH. Such an induced electric field may potentially
lead to various astrophysical effects \citep{Ruffini2023,Levin2018,King2021}.

About half a century ago, Wald investigated the case of a Kerr BH
immersed in an asymptotically uniform magnetic field aligned with
the BH's axis of rotation (where the electromagnetic field is assumed to be too
small to affect the background metric) \citep{Wald1974}. He found
that in this case, a (current-free) four potential $A_{\mu}\left(x\right)$
can be directly constructed from the two Killing fields of the Kerr
background. This construction leads to a two-parameter family of solutions
to Maxwell's equations. The first parameter is the asymptotic magnetic field
magnitude $B_{0}$, and the second parameter, $Q$, is the BH's charge.

An inspection of the $Q=0$ case of Wald's solution for $A_{\mu}\left(x\right)$
shows that along the symmetry axis, there is a non-vanishing, radially
directed, electric field. As a consequence, in an astrophysical environment
that contains free charged particles, the electric field at the pole
will selectively attract charges of one sign over the other, leading
one to anticipate a non-vanishing net charge accretion. In turn, this
net current will cause $Q$ to gradually drift away from zero, eventually
reaching a certain saturation charge $Q_{\mathrm{sat}}\neq0$.

In his paper \citep{Wald1974}, Wald also investigated the expected
value of the saturation charge under the assumption that the so-called
injection energy of charged particles into the BH should vanish at
$Q=Q_{\mathrm{sat}}$. This assumption led to a simple, universal
value for the saturation charge $Q_{\mathrm{sat}}=2B_{0}J\equiv Q_{\mathrm{w}}$
where $J$ denotes the BH's angular momentum. As a technical remark,
note that $Q_{\mathrm{w}}$ is positive if the BH's angular momentum and
the asymptotic magnetic field are co-directed, and negative otherwise.
Throughout this paper, without loss of generality, we shall assume
that $Q_{\mathrm{w}}>0$. Correspondingly, we shall often refer to negatively charged
particles as \emph{attracted }and to positively charged ones as \emph{repelled}, which will be denoted with superscripts ``-" and ``+" respectively.

We are not convinced, however, that the concept of injection energy
can dictate the actual value of the saturation charge. Obviously,
the saturation charge should be determined by the absorption rates for both the positively and negatively charged particles.
To our understanding, the phenomenon of absorption of a particle by
a BH (which is key to the determination of the absorption cross sections and hence, absorption rates)
is a dynamical process, not one governed by equilibrium considerations.

The problem of charge accretion into a magnetized Kerr BH has previously been considered in \cite{Rueda2024} (under the simplifying assumption that charged particles move along the magnetic field lines, even in the BH’s vicinity). Their results seem to differ from Wald’s picture in several respects. In particular, the authors contemplate the possibility that the BH charge oscillates rather than approaches a steady state. Another question mark regarding Wald’s picture was raised in \cite{Li2000} where it was suggested that the system’s total electromagnetic-field energy may be a more relevant criterion than the injection energy of individual charges. The charge that minimizes the total electromagnetic energy is found there to differ from Wald’s charge. In addition, although the system considered in \cite{Adari2023} differs from the one analyzed by Wald (they consider a boosted magnetized BH), we find it relevant to mention the authors’ viewpoint that detailed particle dynamics (rather than mere energetic arguments) should determine the BH’s charge.

Motivated by the desire to better understand the saturation charge
$Q_{\mathrm{sat}}$, in this work we set out to develop an accretion model as natural
and simple as possible, which will allow us to quantitatively explore
the absorption rates in terms of the corresponding cross sections -- and thereby, to determine whether
there actually is charge accretion balance or imbalance at $Q=Q_{\mathrm{w}}$.
In our setup, there are two twin uniform fluxes of unbound positively
and negatively charged particles coming from infinity along the magnetic
field lines. These two fluxes are fully equivalent, in the sense that
they share the same flux density, and the positive and negative charges
have the same mass $m$, the same charge magnitude $\left|e\right|$,
and the same asymptotic incoming velocity -- they only differ by
the sign of the particles' charge (a more detailed description of
our accretion model is given in the next section). We assume that
the fluxes are very dilute, such that individual charges do not affect
each other's motion. In this situation, assuming equal fluxes, the issue of charge accretion
balance (or imbalance) boils down to the question of whether the two
absorption cross sections are equal or not -- which is the question
on which we focus throughout this paper.

Due to the symmetries of the electromagnetic field (and the Kerr metric),
the particles' orbits admit two constants of motion: the energy $E$
and the azimuthal angular momentum $L$. The energy $E$ ($>1$) is
assumed to be the same for all particles composing the two fluxes.
The angular momentum $L$ is uniquely determined by the orbit's impact
parameter $b$. Due to the presence of the electromagnetic field,
Carter's constant is \emph{not} conserved, and as a consequence, we
are unable to solve the equations of motion (EOM) analytically. We
therefore solve them numerically, with the aim of determining whether
a given orbit falls into the BH or not. Our goal is to figure out,
for each of the two charges (for a given set of the model's parameters
$M,J,B_{0},E$ and the particles' charge-to-mass ratio $\frac{e}{m}$),
what is the domain of absorption on the $b$ axis -- which, in turn,
determines the absorption cross section $\sigma$.

Naively, one might expect to find a single continuous domain of absorption
that extends from $b=0$ up to a certain critical value $\tilde{b}$
beyond which no orbit would be absorbed (in which case the cross section
would be $\pi\tilde{b}^{2}$). Indeed, there
is a central, continuous, absorption domain from $b=0$ up to a certain
value $b_{1}$; but to our surprise, we found that at $b>b_{1}$,
the family of orbits has a fractal structure. In particular, in the
range $b>b_{1}$, there appear to exist an infinite set of individual
absorption (as well as non-absorption) domains. We note that such
a fractal behavior has been observed in magnetized Schwarzschild \citep{Frolov2013}
and Kerr \citep{Levin2018,Adari2023} BHs. nevertheless, this fractal
behavior has not been observed or analyzed so far in the context of
a family of unbound orbits coming from infinity.

This fractal behavior makes it extremely challenging to precisely quantify the
absorption cross sections. We therefore opted for a method of evaluating
lower and upper bounds on these cross sections as we now explain.
First, as mentioned above, there always exists a central absorption
domain $0\leq b<b_{1}$. Hence, by numerically finding this parameter
$b_{1}$, we obtain a lower bound on the accretion cross section $\sigma_{\mathrm{min}}=\pi\left(b_{1}\right)^{2}$.
Second, with the help of the constants of motion, we construct an
analytical criterion which tells us in which regions on the $r-\theta$
plane a particle can be present as a function of $b$ (which determines
$L$, as mentioned above). By analyzing this criterion, we obtain
a certain value $b_{0}$ beyond which particles cannot reach the event
horizon at all. This, in turn, gives us an upper bound on the cross
section $\sigma_{\mathrm{max}}=\pi\left(b_{0}\right)^{2}$.

By comparing the lower bound for the attracted charge $\sigma_{\mathrm{min}}^{-}$
with the upper bound for the repelled one $\sigma_{\mathrm{max}}^{+}$,
we find that for all values of BH spin and mass and every $E>1$,
at strong enough magnetic fields, $\sigma_{\mathrm{min}}^{-}>\sigma_{\mathrm{max}}^{+}$.
This indicates that there is net accretion imbalance at $Q=Q_{\mathrm{w}}$,
as the absorption rate of attracted charges is larger than that of
the repelled ones. In turn, this implies that the actual saturation
charge must be smaller than $Q_{\mathrm{w}}$. In fact, we find that
at the limit of strong magnetic field, while $\sigma_{\mathrm{min}}^{-}$
approaches a constant value of order a few times $M^2$, $\sigma_{\mathrm{max}}^{+}$
approaches zero.

In order to investigate charge accretion imbalance, we proceed in
two stages. We start by choosing a specific point in the space of
our model's parameters -- namely specific values of $\frac{J}{M^{2}}$,
$E$ and $B_{0}$ -- for which the analysis becomes especially simple,
and demonstrate imbalance in this specific case. Then, in the next
stage, we extend the analysis to the entire range of $B_{0}$ and
show that charge accretion imbalance is guaranteed to occur beyond
a certain value of $B_{0}$. Moreover, the ``imbalance ratio" $\frac{\sigma^{-}}{\sigma^{+}}$ diverges
at the limit of large $B_{0}$\footnote{As a matter of fact, the magnitude of the magnetic field will later be quantified by means of the dimensionless parameter $\varepsilon$ defined below, rather than $B_0$ itself.}.

In Sec.~\ref{chap:Preliminaries} we review Wald's solution for $A_{\mu}\left(x\right)$
and use it in our accretion analysis. As already mentioned above,
in our model a charged particle comes in from infinity along the magnetic
field lines with a certain energy $E>1$ and impact parameter $b$.
Due to the scale invariance of general relativity, the relevant model's parameters
reduce to the BH's specific spin $\alpha\equiv\frac{J}{M^{2}}$, the
magnetic field strength parameter $\varepsilon\equiv\frac{e}{m}B_{0}M$,
as well as $E$ and $b$. We write down the particles' EOM, and relate
the impact parameter $b$ to the conserved angular momentum $L$.

In Sec.~\ref{chap:Trajectory Space} we investigate the various types
of orbits and explore the structure of the trajectory space, by numerically
solving the EOM for a variety of initial conditions. 
We distinguish between three types of orbits: those falling into the
BH, those escaping to infinity, and orbits that stay trapped forever
in the BH exterior. In our analysis, we focus on the central absorption
domain $0\leq b<b_{1}$ and numerically evaluate $b_{1}$ for a variety
of parameters, providing a lower bound on the absorption cross section.
In addition, we carefully take the $\varepsilon\to-\infty$ limit
of the EOM, which exhibits significant simplification and thereby
provides a valuable insight on the accretion behavior at large magnetic
fields.

In Sec.~\ref{chap:Energy criterion} we implement a complementary
analytical approach based on the constants of motion $E$ and $L$
(as well as the normalization of the four-velocity). This method provides
a simple analytical criterion telling us in which regions in the $r-\theta$
plane the particle is allowed to be present. For each value of $b$
this criterion 
enables us to say whether or not the particle is allowed to be absorbed
into the BH. In turn, this gives us an upper bound $b_{0}$ on the
absorption range on the $b$ axis -- and hence an upper bound on
the absorption cross section.
To facilitate this analysis, we introduce the notion of \emph{critical points} -- points at which the topology of the allowed region changes -- and show how their properties can be used to determine $b_0$. 
For
the repelled particle, our analysis also reveals the presence of an
analytical upper bound $b_{h}\propto\varepsilon^{-\frac{1}{2}}$ on
$b_{0}$.

In Sec.~\ref{chap:Results and analysis} we compare the upper bound
$\sigma_{\mathrm{max}}^{+}=\pi\left(b_{0}^{+}\right)^{2}$ for repelled
particles obtained using the conserved quantities based criterion
with the numerically obtained lower bound $\sigma_{\mathrm{min}}^{-}=\pi\left(b_{1}^{-}\right)^{2}$
for attracted particles, and we demonstrate an imbalance in charge
accretion.

In Sec.~\ref{chap:Discussion} we discuss our results and their physical
implications. In particular, the charge accretion imbalance demonstrated
in Sec.~\ref{chap:Results and analysis} indicates that Wald's charge
$Q_{\mathrm{w}}$ cannot be a universally valid saturation charge
of a magnetized Kerr BH. We briefly mention our preliminary results
for the anticipated correction to Wald's charge, that will be the
focus of a forthcoming paper \citep{Okun2025}. In particular, we
find that Wald's charge remains the correct leading-order saturation
charge in the limit of large $\left|\varepsilon\right|$. We also
demonstrate that this case of large $\left|\varepsilon\right|$ is
astrophysically relevant.

Rotating BHs in magnetized environments are believed to play a central
role in a variety of observed astrophysical phenomena. The results
in this paper shed new light on their charging mechanism (although
only for situations of extremely dilute environments), and in particular
refine our understanding of their steady-state charge. Throughout
this paper we use relativistic units $G=c=1$.

\section{Preliminaries}

\label{chap:Preliminaries} 

\subsection{Wald solution}

\label{sec:Wald solution} It has been shown by Wald \citep{Wald1974}
that the source-free vector potential induced by a stationary and axisymmetric
Kerr BH, placed in an asymptotically uniform magnetic field which
is asymptotically aligned with the axis of rotation, can be constructed
using the timelike and axial Killing vectors $\eta^{\mu}\equiv g_{t}^{\mu}$
and $\psi^{\mu}\equiv g_{\phi}^{\mu}$ respectively, where $g_{\mu\nu}$
is the spacetime metric, and $t$ and $\phi$ are respectively the standard temporal and azimuthal Boyer-Lindquist coordinates, see Eq.~\eqref{eq:kerr metric}. Wald's potential is given by 
\begin{align}
A_{\mu}=\left(\frac{B_{0}J}{M}-\frac{Q}{2M}\right)g_{t\mu}+\frac{1}{2}B_{0}g_{\phi\mu}~,\label{eq:wald potential}
\end{align}
where $M$ and $J$ are respectively the mass and angular momentum
of the BH, and $B_{0}$ is the magnetic field magnitude at infinity.
The parameter $Q$ is the charge of the BH, assumed to be too small
to have an effect on the metric\footnote{The effect of the electromagnetic field (associated with the BH's
charge $Q$) on the metric is expected to be negligible for known
astrophysical examples (see \citep{Wald1974}).}.

It is evident from \eqref{eq:wald potential} that a nontrivial electric
potential is created in the vicinity of the BH. If the BH is surrounded
by charged particles (or plasma), as is often the case in various
astrophysical scenarios, the existence of this electric potential
suggests that selective accretion of charge will take place, leading
to a non-vanishing net charge accretion. This net current gradually
modifies the BH's charge $Q$ until the latter arrives at a \emph{saturation
value} at which the net charge accretion vanishes. Wald suggested
that this saturation charge takes the universal value $Q_{\mathrm{w}}=2B_{0}J$.
This is the special charge value at which the electric field vanishes
everywhere along the symmetry axis $\theta=0$ (leading to a vanishing
injection energy \citep{Wald1974}).

In this paper we analyze the proposed steady state situation $Q=Q_{\mathrm{w}}$
in more detail by analyzing the rate of charge accretion for charges
of both signs, considering the full EOM of a charged test particle
in such a system. It is important to note that this model is applicable
only in situations where the interactions between infalling charges
are negligible. Otherwise, the system may transition from vacuum to
plasma dynamics, and the present analysis is no longer applicable
\citep{Levin2018}. We then enter the realm of magnetohydrodynamics,
which will not be considered in this paper.

In this paper, unless explicitly stated otherwise, we will choose our
units such that the mass of the BH, $M$, is equal to $1$ (in addition
to $G=c=1$). This allows us to rewrite the four-potential as

\begin{align}
\frac{e}{m}A_{\mu}=\varepsilon\left(\alpha\left(1-\lambda\right)\left(g_{t\mu}+\delta_{t\mu}\right)+\frac{1}{2}g_{\phi\mu}\right)~,\label{eq:unitless wald potential-2}
\end{align}
where $\delta_{t\mu}$ denotes Kronecker's delta, $m$ and $e$ are
the particle's mass and charge respectively, and we will refer to
$\varepsilon\equiv\frac{e}{m}B_{0}M$ as the \emph{dimensionless}
\emph{magnetic field strength}, $\alpha\equiv a/M$ is the \emph{dimensionless
spin parameter}, and $\lambda\equiv Q/Q_{\mathrm{w}}$ is the charge
in units of $Q_{\mathrm{w}}$ (We have chosen to present these quantities
with a reintroduction of the BH mass to highlight that they remain
dimensionless regardless of the choice of $M=1$ units). In \eqref{eq:unitless wald potential-2},
we added the gauge term $\varepsilon\alpha\left(1-\lambda\right)\delta_{t\mu}$
such that the electric potential vanishes at infinity. Note that $\mathrm{sign}(\varepsilon)$
depends on the charge of the infalling particle relative to the sign
of $Q_{\mathrm{w}}$. Correspondingly, we refer to the cases of $\mathrm{sign}(\varepsilon)=+1,-1$
as \emph{repelled}/\emph{attracted} particles, respectively. The field
strength is therefore 
\begin{align}
\frac{e}{m}F_{\mu\nu}=\varepsilon\left[\alpha\left(1-\lambda\right)\left(g_{t\nu,\mu}-g_{t\mu,\nu}\right)+\frac{1}{2}\left(g_{\phi\nu,\mu}-g_{\phi\mu,\nu}\right)\right]~.\nonumber \\
\end{align}
In order to study charge accretion at $Q=Q_{\mathrm{w}}$, we will
set $\lambda=1$ unless stated otherwise.

\subsection{Equations of motion}

\label{sec:Equations of motion}

The EOM of a charged particle moving in an electromagnetic field are
\begin{gather}
\frac{du^{\alpha}}{d\tau}+\Gamma_{\mu\nu}^{\alpha}u^{\mu}u^{\nu}=\frac{e}{m}F^{\alpha}{}_{\nu}u^{\nu}\,,\label{eq:geodesic equation}
\end{gather}
where the four-velocity $u^{\mu}$ is also subject to the constraint
$u^{\mu}u_{\mu}=-1$. Note that in this paper, we will not consider
any self-force effects\footnote{ Nevertheless, as will be mentioned below (see Subsec.~\ref{sec:Defining the system}),
in our model we disregard any initial Larmor motion of incoming particles,
and our motivation for this assumption is the radiative decay of any
such cyclotron motion already before the particle has arrived in the
vicinity of the BH.}.

Throughout this paper we assume that the particle is moving in the
Kerr metric, given in Boyer-Lindquist coordinates (with $M$ set to 1) by
\begin{align}
\label{eq:kerr metric}
ds^{2} & =-\left(1-\frac{2r}{\Sigma}\right)dt^{2}-\frac{4r\alpha\sin^{2}\theta}{\Sigma}dtd\phi+\frac{\Sigma}{\Delta}dr^{2}+\Sigma d\theta^{2}\\
 & +\frac{\left(r^{2}+\alpha^{2}\right)^{2}-\Delta\alpha^{2}\sin^{2}\theta}{\Sigma}\sin^{2}\theta d\phi^{2},\nonumber  
\end{align}
where

\begin{align}
\label{eq:metric functions}
\Sigma=r^{2}+\alpha^{2}\cos^{2}\theta\,,\ \ \Delta=r^{2}-2r+\alpha^{2}\,.
\end{align}
The event horizon is located at the $r$ value $r_{+}\equiv1+\sqrt{1-\alpha^{2}}$,
which solves $\Delta=0$.

Since both the metric and the electromagnetic potential are stationary
and axisymmetric, there are two conserved quantities

\begin{equation}
E=-u_{t}-\frac{e}{m}A_{t}~,\ \ L=u_{\phi}+\frac{e}{m}A_{\phi}~,\label{eq:}
\end{equation}
where $E$ and $L$ are respectively the energy and angular momentum
of the particle per unit mass. These two quantities correspond to
the $t$ and $\phi$ components of the four-momentum of a charged
particle in an electromagnetic field:

\begin{align}
\frac{p_{\mu}}{m}=u_{\mu}+\frac{e}{m}A_{\mu}\,.\label{eq:four momentum}
\end{align}

In this paper, we focus on the case of a BH set at Wald's charge,
and correspondingly we take \eqref{eq:unitless wald potential-2}
with $\lambda=1$ as the four-potential. As a consequence, the four-potential
becomes

\begin{align}
\frac{e}{m}A_{\mu}=\frac{1}{2}\varepsilon g_{\phi\mu}~,\label{eq:unitless wald potential}
\end{align}
the field strength

\begin{equation}
\frac{e}{m}F_{\mu\nu}=\frac{1}{2}\varepsilon\left(g_{\phi\nu,\mu}-g_{\phi\mu,\nu}\right)\,,\label{eq:wald charge field}
\end{equation}
and the constants of motion $E$ and $L$

\begin{gather}
E=-u_{t}-\frac{1}{2}\varepsilon g_{\phi t}\,,\label{eq:conserved quantities 1}\\
L=u_{\phi}+\frac{1}{2}\varepsilon g_{\phi\phi}\,.\label{eq:conserved quantities 2}
\end{gather}

Due to the nonlinear nature of the EOM and the lack of a sufficient
number of conserved quantities\footnote{The electromagnetic field \eqref{eq:wald potential} breaks the symmetry
responsible for the conservation of the Carter constant in Kerr \citep{Sun2021}.}, for generic orbital parameters these equations are analytically
intractable, and a numerical approach must be implemented for their
solution. Alternatively, one may use (semi-)analytical energy considerations.
We shall use both of these approaches to derive lower and upper bounds
on the absorption cross section for charged particles.

\subsection{Defining the physical system}

\label{sec:Defining the system}

We consider the motion of test charged particles in the background
of a Kerr BH immersed in an asymptotically uniform magnetic field,
with $A_{\mu}$ corresponding to a Wald-charged BH, given by (\ref{eq:unitless wald potential}).
We assume that all test charges come in from infinity. Far from the
BH region, the particles are moving in a uniform magnetic field. The
most general motion of a charged particle in such a uniform magnetic
field proceeds along a certain fixed magnetic field line, with a cyclotron
motion\emph{ }(also referred to as \emph{Larmor motion}) characterized
by a constant Larmor radius\emph{.} Our model assumes that the incoming
charges arrive from infinity with\emph{ zero Larmor radius}.\emph{
}The physical motivation for this assumption is simple:\emph{ }Since
the charges arrive from infinity, any initial Larmor motion would
be lost due to cyclotron radiation long before the charges approach
the vicinity of the BH.

Each particle's trajectory is characterized by an impact parameter
$b$ (with respect to the symmetry axis $\theta=0$). It is also characterized
by its energy $E>1$, which corresponds to a particle coming from
infinity with asymptotic velocity $>0$. For given initial parameters
$E$ and $b$, we ask whether the particle is absorbed into the BH.

All particles are assumed to be of the same mass $m$ and the same
charge magnitude which is either $+e$ or $-e$. The (asymptotically uniform) influx
of charges of both signs is assumed to be equal, and all particles
have the same energy $E$. We are primarily interested in the absorption
cross sections $\sigma^{\pm}$ for charges of both signs. These two
cross sections depend on the trajectory parameter $E$, as well as
the background parameters $\alpha$ and $\varepsilon$. The overall
charge accretion will be balanced if and only if $\sigma^{+}=\sigma^{-}$.

Since the incoming particles are assumed to have zero initial Larmor
radius, they are asymptotically locked to the magnetic field lines
(which are straight lines parallel to the symmetry axis). Thus, at
the limit in which the incoming particle is still far away from the
BH, its motion is characterized by $r\sin\theta\to const\equiv b$
as well as $\phi\to const$, and the latter also implies $u^{\phi}\to0$.
Far away from the BH we also have $g_{t\phi}\to0$ and $g_{\phi\phi}\rightarrow r^{2}\sin^{2}\theta\rightarrow b^{2}$,
therefore 
\[
u_{\phi}=g_{\phi\phi}u^{\phi}+g_{t\phi}u^{t}\rightarrow b^{2}u^{\phi}\rightarrow0\,.
\]
Substituting these initial far-distance limits of $u_{\phi}$ and
$g_{\phi\phi}$ in Eq. \eqref{eq:conserved quantities 2}, we obtain
the relation between the impact parameter $b$ and the conserved quantity
$L$:

\begin{equation}
L=\frac{1}{2}\varepsilon b^{2}\,.\label{eq:angular momentum}
\end{equation}

We emphasize that throughout this paper, we exclude the trivial case
of $\varepsilon=0$ from our analysis as it does not correspond to
a magnetized BH.

\section{Trajectory space}
\label{chap:Trajectory Space}

The nonlinearity of the EOM \eqref{eq:geodesic equation} and the
system's non-integrable nature (due to the lack of sufficient constants
of motion), imply that trajectories may display a rather complex,
fractal behavior. This behavior becomes evident when numerically solving
the EOM for a variety of initial conditions (see also the discussion
in Sec.~\ref{chap:Discussion}). In particular, the set of absorbed
trajectories as a function of the impact parameter $b$ seems to be
fractal, including an infinite number of absorption and non-absorption
bands. This creates an inherent difficulty in the evaluation of the
particle's absorption cross section. To overcome this complication,
we shall seek for upper and lower bounds on the cross section (for
both attracted and repelled charges), as we further explain below.

\subsection{Trajectory types}

\label{sec:Trajectory types}

For the purpose of investigating the cross section for absorption,
it is useful to categorize particle trajectories according to their
ultimate end point. This results in the definition of three different
trajectory types: 1) \emph{Falling trajectories}, which eventually
cross the event horizon. For example, since the electric field along
the symmetry axis vanishes ($F_{tr}=0$), an incoming charged particle
along the axis of symmetry will necessarily fall into the BH. 2) \emph{Escaping
trajectories}, that eventually go to infinity. For example, a particle
with a large enough impact parameter will inevitably escape to infinity,
only weakly interacting with the BH and the electric potential. 3)
\emph{Trapped trajectories}, that neither fall into the BH nor escape
to infinity. For example, a particle moving in an equatorial circular
orbit. The existence and categorization of such trajectories have
been explored in previous studies \citep{Liu2018,Sun2021,Zaharani2014,Kopacek2018,Aliev2002,Frolov2010}.
In Subsec.~\ref{sec:Critical trajectory} we discuss another type
of trapped trajectory which is of importance to the analysis of domains
of accretion.

The fact that the $b=0$ orbit is always absorbed, combined with the basic
continuity of the solutions to the EOM with respect to the initial conditions,
imply that in some neighborhood of $b=0$ all trajectories are falling.
On the other hand, for large enough $b$ (with $\varepsilon$, $E$
and $\alpha$ remaining fixed), all trajectories are obviously escaping.
We accordingly define $b_{1}$ as the maximal $b$ value up to which
all trajectories are falling, and $b_{2}$ as the minimal $b$ value
beyond which all trajectories are escaping. In the range $b_{1}<b<b_{2}$
trajectories can either fall, escape, or be trapped. We further discuss this complex behaviour (with specific regard to the fractal domain of falling orbits) in Sec.~\ref{chap:Discussion}.

\subsection{\texorpdfstring{Critical trajectory - $b_{1}$}{Critical trajectory
- b1}}

\label{sec:Critical trajectory}

Since, by definition, all particles with $b<b_{1}$ fall into the
BH, we will employ $b_{1}$ to construct a lower bound $\sigma_{\mathrm{min}}$
on the absorption cross section, namely the area of an asymptotic
disk of that radius:

\begin{equation}
\sigma_{\mathrm{min}}\equiv\pi\left(b_{1}\right)^{2}\leq\sigma\,.\label{eq:lower bound b1}
\end{equation}
The trajectory with $b=b_{1}$ is the non-falling trajectory with
minimal $b$ (for given $E,\alpha$ and $\varepsilon$), and we will
refer to it as the \emph{critical trajectory}.

While the definition of $b_{1}$ is simple, it turns out that the
nature of this trajectory is rather subtle. Although not necessary for our analysis,
we give below a brief description of some properties of this critical orbit -- and how
they can help pinpoint it.

The numerical determination of $b_{1}$ may be carried out quite straightforwardly
by scanning along the $b$ axis with high enough resolution until
a non-falling orbit is first encountered. However, the unique properties
of this orbit may be used to identify it in a more efficient way, as we now describe: Fixing $E$, $\alpha$, and $\varepsilon$,
we compute trajectories with gradually increasing $b$ values starting
from $b=0$. For sufficiently small $b$ values, all trajectories
fall into the BH with negative $\dot{r}$ throughout (outside the
BH) -- until one encounters the first falling trajectory where $\dot{r}$
vanishes and subsequently changes sign at some proper time $\tau$.
We will refer to such points of sign change as \emph{bounces}. The
number of bounces steadily grows as we further increase $b$ towards
$b_{1}$. In the limit $b\to b_{1}$, the number of bounces tends
towards infinity -- which designates the critical trajectory $b_{1}$.

The critical trajectory $b=b_{1}$ itself is thus characterized by an oscillatory
motion -- which eventually asymptotes to a periodic orbit in both
the $r$ and $\theta$ coordinates. This also translates to an asymptotically
periodic oscillatory motion in the $\rho-z$ plane, where $\rho$
and $z$ are cylindrical coordinates defined using the Boyer-Lindquist
coordinates as 
\begin{gather}
\rho=\sqrt{r^{2}+\alpha^{2}}\sin\theta\,,\label{eq:cylindrical coordinates}\\
z=r\cos\theta\,.
\end{gather}
We further elaborate on these properties of the $b=b_1$ orbit in App.~\ref{app:critical trajectory}.

While for $b<b_{1}$ the behavior is simple in the sense that all
trajectories are of the falling type, the situation is drastically
different for $b>b_{1}$, where the set of trajectories appears to
become fractal (with regards to their types). In particular, any neighborhood
$b>b_{1}$ of $b_{1}$, no matter how small, seems to contain both
falling and escaping orbits (as well as trapped ones in between).
In fact, to our understanding, within any such neighborhood, there
exists an infinite sequence of alternating segments of falling and
escaping orbits.

\subsection{Universal limit of large $\varepsilon$}

\label{sec:Universal limit}

The limit of strong magnetic field for attracted charged particles,
$\varepsilon\to-\infty$, is inherently, as well as astrophysically
(see Sec.~\ref{chap:Discussion}), interesting. In that limit, an
attracted particle with $b>0$ moving towards the BH will be accelerated
by the electric field to ultra-relativistic velocities (as opposed
to a repelled particle, that would be slowed down). Since the electric
field is linear in $\varepsilon$, the charge will become ultra-relativistic
already in the weak field region. That is, at \emph{any} generic fixed
point in space, the four-velocity $u^{\mu}$ diverges like $\varepsilon$,
hence the particle's trajectory will asymptotically tend to a light-like
curve.

It is convenient to investigate this limit at the level of the EOM
by reparametrizing the orbit with $\tau'=\varepsilon\tau$ instead
of proper time $\tau$, and correspondingly, defining the rescaled
four-velocity $u'^{\mu}=\frac{dx^{\mu}}{d\tau'}$. Importantly, substituting
these new variables in Eq.~(\ref{eq:geodesic equation}) yields well-behaved
and effective EOM, as will be shown below. We refer to these EOM as
``universal'' as they are independent of both $\varepsilon$ and
$E$, as we will now discuss.

Under the reparametrization, the rescaled four-velocity normalization
condition becomes 
\begin{align*}
u'^{\mu}u'_{\mu} & =-\frac{1}{\varepsilon^{2}}\:\underset{\varepsilon\to-\infty}{\longrightarrow}0.
\end{align*}

Also, replacing $u_{\mu}\to\varepsilon u'_{\mu}$ in Eqs.~\eqref{eq:conserved quantities 1},
\eqref{eq:conserved quantities 2} and dividing by $\varepsilon$
reveals that $E'=E/\varepsilon$, $L'=L/\varepsilon$ play a role
analogous to that of the energy and azimuthal angular momentum in
the reparametrized problem. As explained earlier, in our physical
setup we fix $E$ and investigate the orbits as a function of $b$,
for fixed $\varepsilon$. Correspondingly, when we rescale the problem
and take the $\varepsilon\to-\infty$ limit (with the purpose of understanding
the system's behavior at strong magnetic fields), the original parameter
$E$ is kept fixed. This implies that the rescaled energy vanishes:

\begin{align}
E' & =0.\label{eq:universal energy}
\end{align}
For the angular momentum the situation is somewhat different: for
a given impact parameter $b$ we have $L=\frac{1}{2}\varepsilon b^{2}$
as seen in Eq.~(\ref{eq:angular momentum}) which, when rescaled,
becomes 
\begin{align}
L' & =\frac{1}{2}b^{2}.\label{eq:universal angular momentum}
\end{align}
In the universal limit, therefore, there is \emph{only} \emph{one}
\emph{non-trivial constant of motion} $L'$. The fact that in this
limit the dynamics of the particle are independent of the energy (and
obviously $\varepsilon$) is the main reason behind our decision to
refer to this limit as ``universal''.

At the universal limit $\varepsilon\to-\infty$, the EOM take the
form 
\begin{gather}
\frac{du'^{\alpha}}{d\tau'}+\Gamma_{\mu\nu}^{\alpha}u'^{\mu}u'^{\nu}=\tilde{F}^{\alpha}{}_{\nu}u'^{\nu}\,,\label{eq:universal geodesic equation}
\end{gather}
combined with the null normalization condition

\begin{equation}
u'^{\mu}u'_{\mu}=0~,\label{eq:universal normalization-2}
\end{equation}
where $\tilde{F}\equiv F/\varepsilon$ is the rescaled electromagnetic
tensor, which depends only on the spin parameter $\alpha$. Thus,
the EOM in the universal limit portray a picture of a massless charged
particle with zero energy and fixed angular momentum, moving in a
magnetized Kerr background with finite field strength $\tilde{F}$.

\begin{figure}
\centering \includegraphics[width=0.5\textwidth]{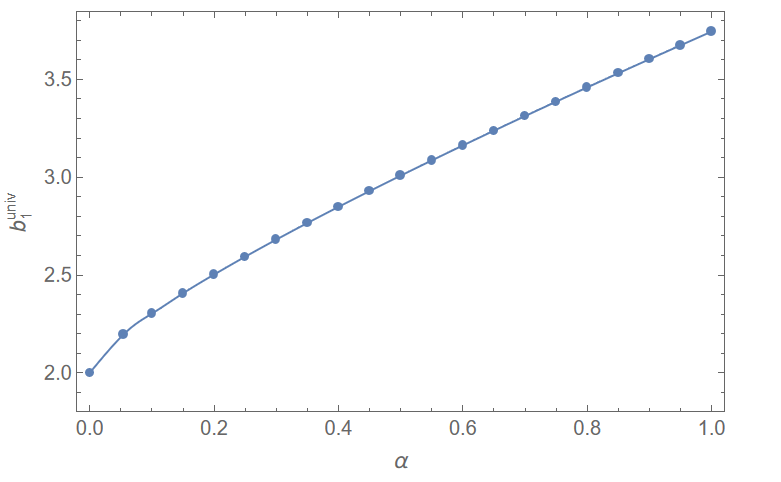}
\caption{The critical impact parameter $b_{1}^{\mathrm{univ}}$ in the universal
limit $\varepsilon\to-\infty$, obtained numerically as a function
of the spin parameter $\alpha$.}
\label{fig:universal critical b1} 
\end{figure}

\begin{figure}
\centering \includegraphics[width=0.5\textwidth]{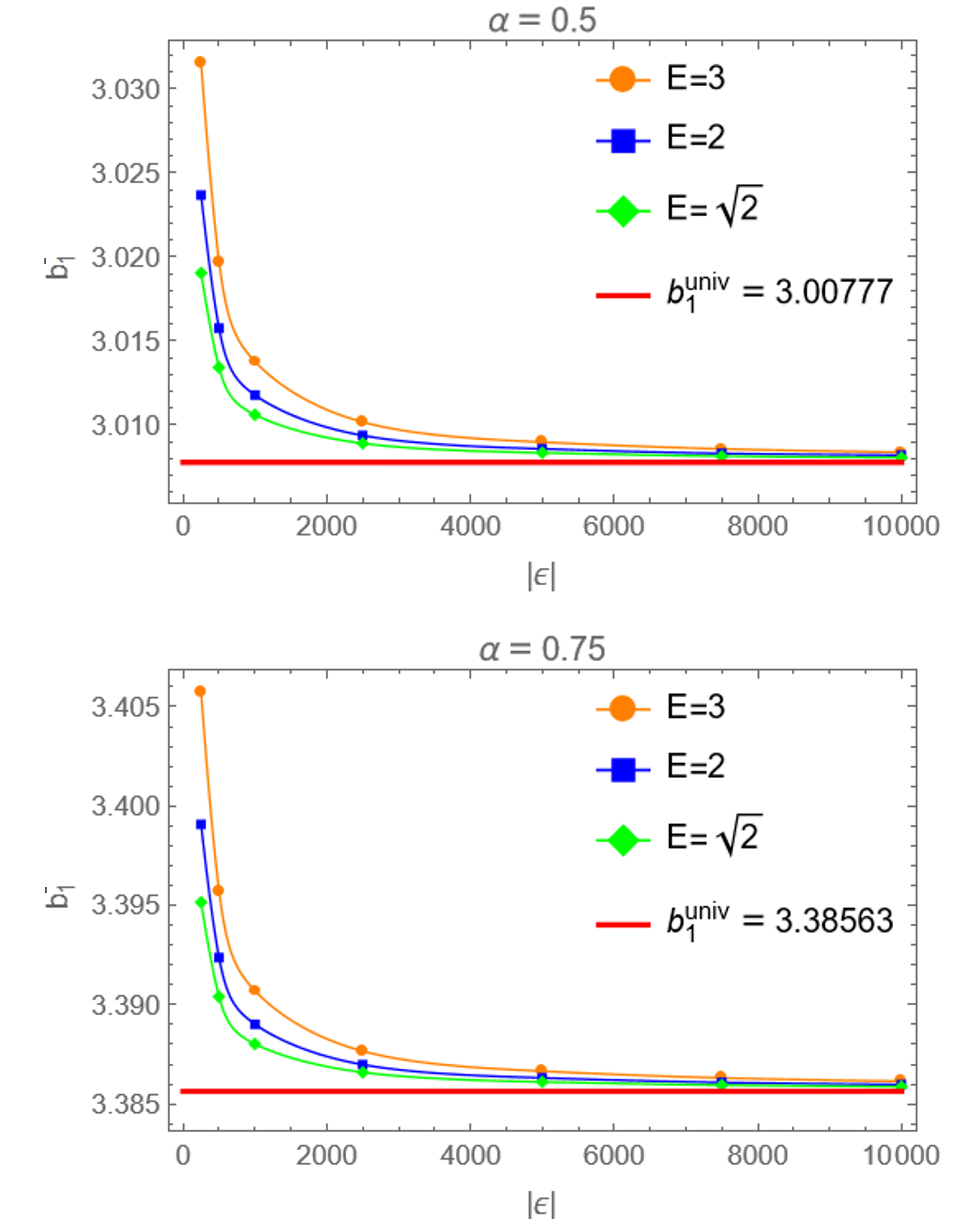}
\caption{Evolution of the critical impact parameter for an attracted particle,
$b_{1}^{-}$, in the large-$\varepsilon$ limit for three initial
energies (different shapes and colors) at two values of the spin parameter
$\alpha$ (different panels). The values of $b_{1}^{-}$ for all initial
energies converge toward the universal limit for the same $\alpha$.}
\label{fig:convergence to universal b1} 
\end{figure}

While the EOM greatly simplify in the universal limit, they are still
analytically intractable. Nevertheless, considering this limit enhances our understanding of the critical
trajectory $b=b_{1}$ at $(-\varepsilon)\gg1$. It also provides important
insights on the (im)possibility of presence of the particle at any
given point in space through the energy considerations described in
Sec.~\ref{chap:Energy criterion} in the $(-\varepsilon)\gg1$ limit;
we elaborate on this in App.~\ref{b0m limit}. The critical trajectory
\[
b_{1}^{\mathrm{univ}}\left(\alpha\right)\equiv\lim_{\varepsilon\to-\infty}b_{1}\left(\varepsilon;\alpha,E\right)
\]
can be found numerically using the same method presented in Subsec.~\ref{sec:Critical trajectory},
and it presents the same qualitative behavior as described therein.
Fig.~\ref{fig:universal critical b1} shows numerical results\footnote{Note that in Fig.~\ref{fig:universal critical b1}, as well as in Figs.~\ref{fig:convergence to universal b1},\ref{fig:main graph} and \ref{fig:imbalance graph}, the actual numerically calculated data are represented by points on the graphs while the curves connecting them are merely interpolations added for better visual illustration.} for
the universal critical impact parameter $b_{1}^{\mathrm{univ}}$ as
a function of spin $\alpha$, computed using the universal EOM (\ref{eq:universal geodesic equation}).
As expected, for any fixed $\alpha$, $b_{1}\left(\varepsilon;\alpha,E\right)$
converges to the corresponding critical impact parameter $b_{1}^{\mathrm{univ}}$
as $\varepsilon\rightarrow-\infty$, and this limiting parameter is
$E$-independent. This is illustrated in Fig.~\ref{fig:convergence to universal b1}. 


\section{Conserved Quantities Based (CQB) Criterion}
\label{chap:Conserved Quantities Based (CQB) Criterion}

\label{chap:Energy criterion}

\subsection{Energetically allowed and forbidden regions}

\label{sec:Allowed regions and critical points}

Instead of directly solving the EOM, one can seek an upper bound on
the value of $b$ for infalling trajectories -- which, in turn, will
provide an upper bound on the absorption cross section for charged
particles. Here we explain how such an upper bound can be found using
the particle's constants of motion.

The timelike normalization requirement $u^{\mu}u_{\mu}=-1$ can be
employed to construct a useful criterion based on the following quantity:
\begin{align}
n\equiv g_{tt}\left(u^{t}\right)^{2}+2g_{t\phi}u^{t}u^{\phi}+g_{\phi\phi}\left(u^{\phi}\right)^{2}+1\,.\label{eq:energy criterion}
\end{align}
We are concerned here with the domain $r>r_{+}$, where both $g_{rr}$
and $g_{\theta\theta}$ are greater than zero. Therefore, $g_{rr}(u^{r})^{2}+g_{\theta\theta}(u^{\theta})^{2}$
is always non-negative, implying that $n$ must be \emph{non-positive}
for timelike orbits. Using the conserved quantities (\ref{eq:conserved quantities 1}),(\ref{eq:conserved quantities 2}),
we express $u^{t}$ and $u^{\phi}$ as functions of $r$ and $\theta$:

\begin{equation}
\begin{aligned}
\label{eq:t and phi components}
& u^{t}=E+r\frac{2E\left(r^{2}+\alpha^{2}\right)-b^{2}\alpha\varepsilon}{\Delta\Sigma},\\
& u^{\phi}=-\frac{1}{2}\varepsilon+\frac{r\alpha\left(2E\left(r^{2}+\alpha^{2}\right)-b^{2}\alpha\varepsilon\right)}{\left(r^{2}+\alpha^{2}\right)\Delta\Sigma} \\
& +\frac{b^{2}\varepsilon}{2\left(r^{2}+\alpha^{2}\right)\sin^{2}\theta}\,.
\end{aligned}
\end{equation}

When plugged into (\ref{eq:energy criterion}), this gives us $n$
as a function of $r$ and $\theta$: 
\begin{align}
n(r,\theta;b,E,\alpha,\varepsilon)=\frac{n_{\mathrm{u}}}{4\sin^{2}\theta\Delta\Sigma}~,\label{eq:fullnoramlization}
\end{align}
where $n_{\mathrm{u}}$ is a polynomial in all its parameters, as
well as in $r$ and $\sin^{2}\theta$, which is given by: 

\begin{equation}
\begin{aligned}
\label{eq:numeratornormalization}
& n_{\mathrm{u}}(r,\theta;b,E,\alpha,\varepsilon)=b^{4}\varepsilon^{2}\Delta \\
& +\sin^{2}\theta\left[4\left(r^{2}+\alpha^{2}\right)\left(\Delta-E^{2}\left(r^{2}+\alpha^{2}\right)\right)+8b^{2}Er\alpha\varepsilon\right] \\
& -\sin^{2}\theta\left[b^{2}\alpha^{2}+2\Delta\left(r^{2}+\alpha^{2}\right)\right]b^{2}\varepsilon^{2} \\
&+\sin^{4}\theta\Delta\left[4\left(E^{2}-1\right)\alpha^{2}+\left(2b^{2}\alpha^{2}+\left(r^{2}+\alpha^{2}\right)^{2}\right)\varepsilon^{2}\right] \\
&-\sin^{6}\theta\varepsilon^{2}\alpha^{2}\Delta^{2}\,.
\end{aligned}
\end{equation}

Note that the function $n$ depends only on the coordinates $r$ and
$\theta$ (in addition to its dependence on the system parameters
$\alpha$ and $\varepsilon$, and the orbit parameters $E$ and $b$).
As mentioned, the particles' orbits must satisfy

\begin{align}
n(r,\theta;b,E,\alpha,\varepsilon)\leq0~.\label{eq:cqb criterion}
\end{align}
We shall refer to this inequality as the \emph{conserved-quantities
based} (CQB) \emph{criterion} (sometimes also referred to as the ``energy
condition'' or ``energy considerations'' etc.).

Given fixed values of $\alpha,\varepsilon$ and $E$, our goal here
is to compute an upper bound on the absorption cross section. The
criterion defined in Eq.~(\ref{eq:cqb criterion}) allows us, for
every choice of $b$, to divide the $r-\theta$ plane into two distinct
regions: (i) The \emph{allowed region}, defined by $n\leq0$, is the
region in which timelike trajectories can energetically be present;
(ii) the \emph{forbidden region}, defined by $n>0$, is the region
in which timelike trajectories cannot exist, as dictated by the CQB
criterion. 
While similar methods have been used in previous studies \citep{Kopacek2018,Gupta2021},
to our knowledge, it has not been used to consider upper bounds on
the absorption cross section of the BH for fluxes of unbound charged
particles coming from infinity.

Obviously the entire trajectory of the particle must be contained
in an allowed region. Therefore, for a particle to be able to fall
into the BH, a connected allowed region that extends from infinity
to the event horizon must exist. This implies that for this CQB analysis
it is sufficient to restrict ourselves to the domain $r>r_{+}$. In
this domain, the denominator in the right hand side of Eq.~(\ref{eq:fullnoramlization})
is always positive (except for the case of $\theta=0$ which will
be dealt with separately below) and the sign of $n$ is the same as
that of $n_{\mathrm{u}}$, therefore the condition in Eq.~(\ref{eq:cqb criterion})
reduces to

\begin{equation}
n_{\mathrm{u}}(r,\theta;b,E,\alpha,\varepsilon)\leq0~.\label{eq:nu criterion}
\end{equation}
Note that this criterion is mathematically more convenient than the
one in Eq.(\ref{eq:cqb criterion}) due to the polynomial nature of
$n_{\mathrm{u}}$.

On the axis of symmetry, namely at $\theta=0$ (or equivalently $\theta=\pi$), the denominator of
$n$ in Eq. (\ref{eq:fullnoramlization}) vanishes. If $n_{\mathrm{u}}>0$
on the axis, then $n$ diverges to $+\infty$ there, meaning that
the axis and its neighborhood are forbidden in such a case. Therefore,
in order for the axis and its neighborhood to be energetically allowed,
Eq. (\ref{eq:nu criterion}) must hold at $\theta=0$ as well.

\subsection{Globally connecting and disconnecting points}

\label{sec:Globally connecting and disconnecting points}

Due to the vanishing of the electric potential along the symmetry
axis, particles with $b=0$ fall directly into the BH. By continuity,
there exists a $b$-range for which all particles are absorbed into
the BH. It therefore follows that for a small enough $b$, all $b$
values are energetically allowed for absorption.

On the other hand, all orbits with sufficiently large $b$ values
are forbidden for absorption. To see this, note that (i) $n_{\mathrm{u}}$
is a polynomial of second order in $b^{2}$ whose coefficients are
themselves polynomial in $r$ and $\sin^{2}\theta$; (ii) the coefficient
of the highest order in $b$ (namely, $b^{4}$) is $\varepsilon^{2}r\left(r-2\right)+\varepsilon^{2}\alpha^{2}\left(1-\sin^{2}\theta\right)$,
which is always \emph{positive} for $r>2$. Choose now any ``circle''
of $r=const>2$ and evaluate the sign of $n_{\mathrm{u}}$ on this
curve. Both the coefficients of the $b^{2}$ and $b^{0}$ terms in
$n_{\mathrm{u}}$, being polynomials in $\sin^{2}\theta$, are bounded
from above at that $r$ value. Since the coefficient of the highest order
term $b^{4}$ is positive, it is obvious that at sufficiently large
$b$,  $n_{\mathrm{u}}>0$ is satisfied everywhere on this constant-$r$ curve, meaning that
this circle is entirely in a forbidden region. In other words, there
is no allowed region connecting the horizon to infinity.

From the above, it follows that there must exist a value of $b$ that
marks a transition (with increasing $b$) from a $b$-range for which absorption is allowed to one for which it is forbidden. We call such a transition point (on the
$b$ axis) a \emph{globally disconnecting point}. In principle, there
may also exist a point of transition in the other direction (namely,
from a forbidden range to an allowed one upon increasing $b$) which we call a \emph{globally
connecting point}.

The globally disconnecting point of largest $b$ is of utmost importance
to the question of accretion imbalance because it provides an upper
bound on the accretion cross section. We denote this point by $b_{0}$
and the bound it provides on the cross section is

\begin{equation}
\sigma\leq\pi\left(b_{0}\right){}^{2}\equiv\sigma_{\mathrm{max}}.\label{eq:upper bound b0}
\end{equation}

Generally speaking, there can be two different situations: 
\begin{enumerate}
\item There is only one globally disconnecting point (and no globally connecting
points), in which case it corresponds to $b_{0}$. 
\item There are multiple globally connecting/disconnecting points, in which
case $b_{0}$ will be the maximal globally disconnecting point. 
\end{enumerate}
We shall refer to the second case (unlike the first one) as having
``energetic rings''. Each such ring may potentially provide a contribution
to the absorption cross section (which nevertheless does not affect
the upper bound $\pi\left(b_{0}\right)^{2}$). As we discuss later,
at least in the cases considered in this paper, there actually are
no such energetic rings.

\subsection{Critical points and their classification}

\label{sec:Critical points}

Finding the global connecting and disconnecting points -- and most
importantly $b_{0}$ -- is not an easy task. In principle, one could
map the allowed and forbidden regions for a dense set of individual
$b$ values, and visually extract from it the approximate $b$ values
of the globally (dis)connecting points. While straightforward, this
method is computationally highly costly. As a better way for this
investigation, we developed the method of \emph{critical points},
which we here describe.

\begin{figure}
\centering \includegraphics[width=0.5\textwidth]{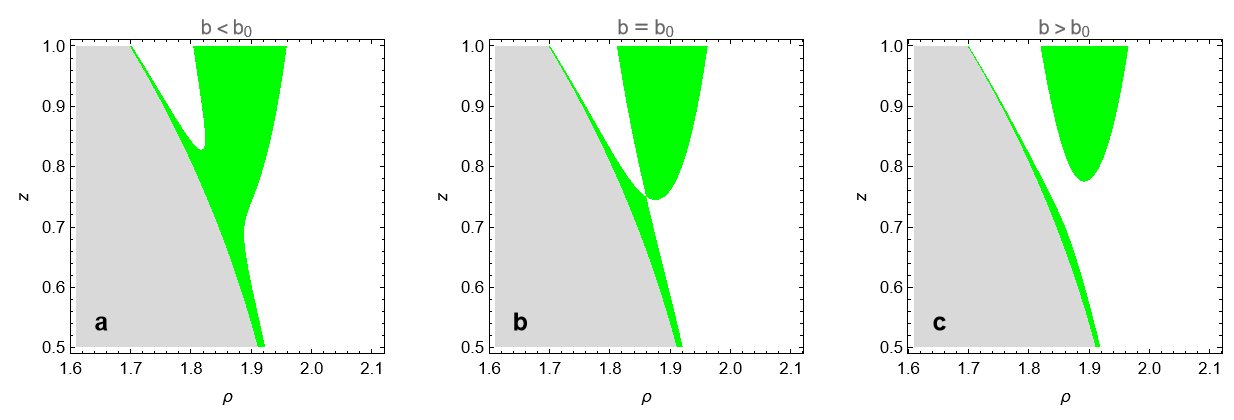}
\caption{Evolution with $b$ of the allowed region for particle presence (green)
for $E=\sqrt{2}$, $\alpha=0.3$, and $\varepsilon=10$. The (non-equatorial)
critical point occurs at $b_{0}=1.87777$. The gray region represents
the region $r<r_{+}$. Panel \textbf{a} shows the allowed region for
$b<b_{0}$ which permits the absorption of the charged particle; panel
\textbf{b} shows the disconnection of the horizon from infinity, precisely
at the critical point $b_{0}$; and panel \textbf{c} shows the disconnected
allowed regions for $b>b_{0}$.}
\label{fig:b0 disconnecting 2D maps} 
\end{figure}

Recalling the continuity of $n_{\mathrm{u}}$ (see Eq. (\ref{eq:numeratornormalization})),
for fixed $E$, $\alpha$, $\varepsilon$, any globally disconnecting
(connecting) point that occurs at some particular $b$ manifests itself
in the $r-\theta$ plane as a point where an allowed (forbidden) region
splits into two distinct regions which pinch off from each other\footnote{Hypothetically, other types of (dis)connections might be possible
for which at the critical $b$ value the (dis)connection occurs along
a curve in the $r-\theta$ plane rather than at a single point. However,
an analytical examination (based on the \emph{resultant} method) reveals
that no such curves exist in the space of solutions.}, separated by a forbidden (allowed) region; see Fig.~\ref{fig:b0 disconnecting 2D maps}
for an example of such a disconnection process at a point $b=b_{0}$.
At each such pinch-off point, not only $n_{\mathrm{u}}$ vanishes
but also its gradient must vanish, therefore, the following set of
equations must be satisfied there: 
\begin{gather}
n_{\mathrm{u}}=0\,,\label{eq:critical equations 1}\\
n_{\mathrm{u},r}=0\,,\label{eq:critical equations 2}\\
n_{\mathrm{u},\theta}=0\,.\label{eq:critical equations 3}
\end{gather}
We conclude that all globally connecting and disconnecting points
must correspond to solutions of this set of equations\footnote{This system of equations can easily be converted to a set of \emph{polynomial}
equations in all its variables, by replacing the variable $\theta$
with $v\equiv\sin^{2}\theta$ -- which is more convenient for analysis.\label{fn:polynomial}}. This fact provides a convenient method for finding the globally
(dis)connecting points by numerically solving this set of equations.
Mathematically, this set of equations typically has numerous solutions
but we restrict ourselves to ``physical'' solutions, which are real
solutions with $r>r_{+},0<\sin^{2}{\theta}\leq1$ and $b>0$. We refer
to these ``physical'' solutions as \emph{critical points}\footnote{Not to be confused with the notion of ``critical trajectory'' which
corresponds to $b=b_{1}$ mentioned in Subsec. \ref{sec:Critical trajectory}.}$^{,}$\footnote{Interestingly, critical points can be shown to fully correspond to
circular trajectories at a (not necessarily equatorial) constant $\theta$.
We hope to elaborate on this elsewhere.}. We empirically find that there always are at least one and at most
four such critical points (for given $0<\alpha<1,\:E>1$ and $\varepsilon\neq0$)\footnote{Practically, we find the critical points by solving the polynomial
variant of this set of equations (see footnote \ref{fn:polynomial}).
To this end, we use \emph{Wolfram Mathematica}'s command \emph{NSolve,
}which solves this algebraic system easily and with any desired precision. }.

Since $n_{\mathrm{u}}$ is a polynomial in $\sin^{2}\theta$, any
point at the equatorial plane automatically satisfies $n_{\mathrm{u},\theta}=0$.
Therefore, the equatorial critical points can be obtained by solving
only the system of two equations $n_{\mathrm{u}}=n_{\mathrm{u},r}=0$
(with $\theta=\frac{\pi}{2}$) for the two unknowns $r$ and $b$.

It is important to recall that not all critical points necessarily
correspond to globally (dis)connecting points. That is, a (dis)connection
of a certain local region does not necessarily affect the global connectivity
between the horizon and infinity. Therefore, a critical point should
merely be regarded as a \emph{candidate} for a globally (dis)connecting
point. Despite this subtlety, the critical points provide a very efficient
method for finding the globally (dis)connecting points: as mentioned
above, we found that there are only up to 4 critical points, and it
is not difficult to figure out which of them are the globally (dis)connecting
points. In fact, local properties of critical points provide effective
information about their ``global candidacy'', as we now explain.

First, a point where the gradient of $n_{\mathrm{u}}$ vanishes (as
well as $n_{\mathrm{u}}$ itself) may be either (i) a local maximum/minimum
or (ii) a saddle-point of $n_{\mathrm{u}}$. Only the second case
is of relevance to us, as case (i) merely represents the appearance/disappearance
of an allowed/forbidden region -- which does not affect the global
connectivity. In other words, we are only interested here in \emph{saddle-type}
critical points. This adds the condition $\det H<0$, where $\det H=n_{\mathrm{u},rr}n_{\mathrm{u},\theta\theta}-\left(n_{\mathrm{u},r\theta}\right)^{2}$
is the Hessian matrix determinant associated with $n_{\mathrm{u}}\left(r,\theta\right)$.

\begin{figure}
\centering \includegraphics[width=0.5\textwidth]{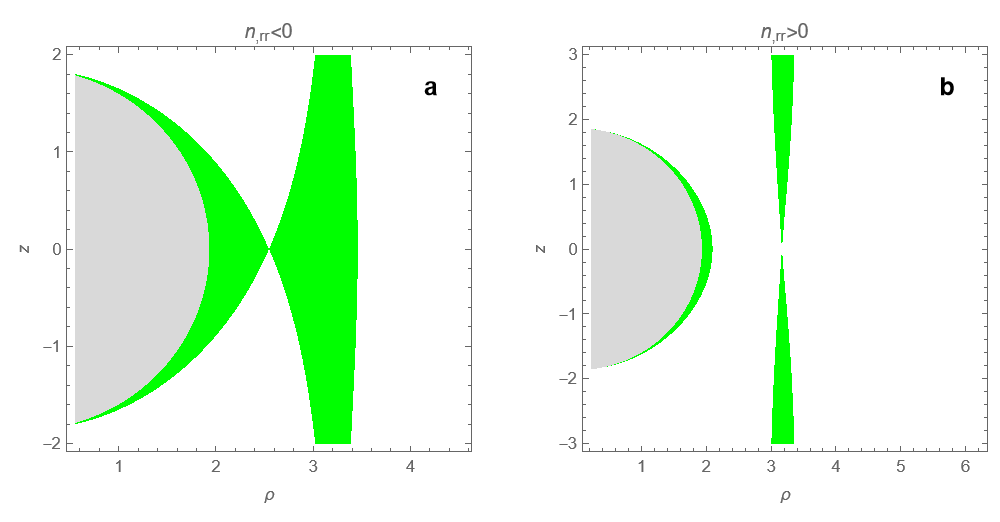}
\caption{Maps of allowed (green) and forbidden (white) regions for particle
presence around a critical point. The gray region around the origin
represents the zone $r<r_{+}$. In panel \textbf{a} the equatorial
critical point has $n_{\mathrm{u},rr}<0$ and can therefore correspond
to a (dis)connection from infinity to the horizon, while in panel
\textbf{b} the critical point has $n_{\mathrm{u},rr}>0$ and cannot
correspond to a (dis)connection from infinity to the horizon. The
parameters for both panels are $E=3,\,\alpha=0.5$, while $\varepsilon=-100$
for panel \textbf{a} and $\varepsilon=15$ for panel \textbf{b}.}
\label{fig:critical point orientation} 
\end{figure}

The second local criterion concerns only equatorial critical points:
since the $\theta$-dependence of $n_{\mathrm{u}}$ is only through
$\sin^{2}\theta$, the allowed and forbidden regions have a reflection
symmetry with respect to the equatorial plane. There may be equatorial
critical points that merely (dis)connect allowed regions from the
two sides of the equatorial plane. Such a (dis)connection has nothing
to do with global connectivity between the event horizon and infinity.
Therefore, we are only interested in equatorial critical points that
connect two allowed regions which both have existence at the ``north''
(as well as the ``south'') side of the equatorial plane (or in other
words -- a critical point that connects two \emph{forbidden} regions
from the two sides of the equatorial plane). This structure is displayed
in panel \textbf{a }of Fig. \ref{fig:critical point orientation}
(unlike the situation in panel \textbf{b}). Such equatorial critical
points satisfy $n_{\mathrm{u},rr}<0$.

Summarizing the classification analysis above, a critical point will
be referred to as a \emph{relevant point} if it satisfies the following
two inequalities:

\begin{align*}
\det H & <0\\
n_{\mathrm{u},rr} & <0\ \ \ \ \ (\text{for equatorial points only})
\end{align*}

Among the relevant points, some satisfy $n_{\mathrm{u},b}>0$, which
implies that two allowed regions locally disconnect as $b$ is increased
at this point in space; see Fig.~\ref{fig:b0 disconnecting 2D maps}.
We refer to such relevant points as \emph{locally disconnecting points}.
Likewise, relevant points at which $n_{\mathrm{u},b}<0$, will be
referred to as \emph{locally connecting points}. Any globally (dis)connecting
point must in particular be a locally (dis)connecting point. Importantly,
note that if there is only one locally disconnecting point, it must
correspond to $b_{0}$.

As mentioned above, in principle there might be ``energetic rings''.
In this case, there must be at least one globally connecting point
-- and also at least two globally disconnecting ones. As a matter
of fact, based on the local analysis of critical points developed
above, we conclude (by a combination of analytical considerations and empirical findings, which we hope to present elsewhere) that there are no locally connecting points for any point in the $E$ and $\alpha$ parameter space. This of
course implies that there are no globally connecting points, and thus,
there are no ``energetic rings''. This means that the entire domain
$b<b_{0}$ is energetically allowed for accretion, while the entire domain $b>b_{0}$ is forbidden.

The scope of this paper includes an exploration of the entire $\varepsilon$
axis (for specific values of $\alpha$ and $E$; see Subsec. \ref{sec:Accretion imbalance as a function of =00003D000024=00003D00005Cvarepsilon=00003D000024})
using the critical point analysis described above. The exploration
of the $\varepsilon$ axis is done in two independent ways: 
\begin{enumerate}
\item Applying the above mentioned critical-points classification on a dense
set of $\varepsilon$ values (for the chosen $\alpha$ and $E$). 
\item Using a more analytical approach, one finds that the $\varepsilon$
axis is divided into a certain number of non-overlapping domains (whose
union covers the entire $\varepsilon$ axis, except $\varepsilon=0$,
which is entirely excluded from our analysis), in each of which the
number of critical points -- and also their classification -- is
uniform. Each pair of adjacent domains is separated by a \emph{transition
point}, which is a certain $\varepsilon$ value that can usually be
obtained by solving a certain set of algebraic equations (which we
hope to present and discuss elsewhere) for the four unknowns $r,\theta,b$
and $\varepsilon$. In the specific case presented in Subsec. \ref{sec:Accretion imbalance as a function of =00003D000024=00003D00005Cvarepsilon=00003D000024},
in the range $\varepsilon>0$, there are four such transition points
which divide the $\varepsilon$ axis into five domains. The qualitative
features derived from the local properties of the critical points,
for example the lack of a local connecting point, cannot change within
a particular domain. Hence, the characteristic features of the critical
points within a certain $\varepsilon$-domain can be deduced by investigating
the critical points at a single, representative, $\varepsilon$ value
in that domain. On the other hand, in the range $\varepsilon<0$ there
are no transition points -- which defines the entire negative-$\varepsilon$
range as a single domain. These conclusions (for both $\varepsilon>0$
and $\varepsilon<0$) are also confirmed by the search over a dense
set of $\varepsilon$ values carried out in method 1. 
\end{enumerate}
Using the latter method we find that in the specific case presented
in Subsec. \ref{sec:Accretion imbalance as a function of =00003D000024=00003D00005Cvarepsilon=00003D000024}
($\alpha=0.75,E=\sqrt{2}$), in the first four domains there is only
a single locally disconnecting point -- which must be $b_{0}$. In
particular, this is the situation for the specific value $\varepsilon=2$
discussed in further details in Subsec.~\ref{sec:Accretion imbalance at =00003D000024Q_=00003D00005Cmathrm=00003D00007Bw=00003D00007D=00003D000024}
(this particular $\varepsilon$ value belongs to the first domain).
In the fifth domain, corresponding to $\varepsilon$ values larger
than the last transition point $\varepsilon_{\mathrm{last}}\approx6.8702$,
there actually are two locally disconnecting points. The identification
of $b_{0}$ in this case is done using a complementary analytical
method as we describe in the next subsection. The implementation of
the classification process for this last $\varepsilon$ domain is
demonstrated in App. \ref{app:classification} and compactly summarized in Table~\ref{table}.

It turns out that for sufficiently large positive $\varepsilon$ there
always is a non-equatorial critical point. In this limit, the critical
point approaches the polar axis, at a limiting value $r\to r_{c}$
slightly larger than $r_{+}$. The corresponding $b$ value approaches
zero as $b\approx c\varepsilon^{-\frac{1}{2}}$, where $c$ is some
positive coefficient that, like $r_{c}$, depends on $E$ and $\alpha$.
The analysis of the asymptotic behavior of the set of equations (\ref{eq:critical equations 1}),(\ref{eq:critical equations 2}),(\ref{eq:critical equations 3})
at $\varepsilon\gg1$ -- which is out of the scope of this paper
-- provides the values of both $r_{c}\left(\alpha,E\right)$ and
$c\left(\alpha,E\right)$ (by solving a certain polynomial equation).
A systematic check of a large dense set of $\alpha$ and $E$ values
indicates that there exist only a single such critical point at large
$\varepsilon$, which always corresponds to $b_{0}$.

\subsection{An analytical upper bound}

\label{sec:An analytical upper bound}

While investigating the properties of $n_{\mathrm{u}}$ we found,
to our surprise, that for $\varepsilon>0$ there exists a special
value of $b$ for which $n_{\mathrm{u}}$ vanishes on the event horizon
for all $\theta$. To show this, we substitute $r=r_{+}$ in Eq. (\ref{eq:numeratornormalization})
which yields 
\begin{gather}
n_{u}(r_{+},\theta)=-\sin^{2}\theta\left[b^{4}\alpha^{2}\varepsilon^{2}-8b^{2}Er_{+}\alpha\varepsilon+16r_{+}^{2}E^{2}\right]\,.\label{eq:numeratorathorizon}
\end{gather}
Notice that the term in the brackets is independent of $\theta$.
Equating this expression to zero and solving for $b$ gives a single
real solution in the relevant range $b>0$: 
\begin{align}
b=\sqrt{\frac{4Er_{+}}{\alpha\varepsilon}}\equiv b_{h}\,,\label{eq:bh}
\end{align}
Considering first the situation at $b=b_{h}$, in App.~\ref{app:nrposdem}
we show that at $r=r_{+}$, $n_{\mathrm{u},r}>0$ is satisfied for all $\theta$
(including $\theta=0$), which implies that $n_{\mathrm{u}}>0$ throughout some $r>r_{+}$ neighborhood of the event horizon. In other words, particles
coming towards the BH with $b=b_{h}$ cannot be absorbed as their
presence near the event horizon is forbidden. Because $n_{\mathrm{u}}$
is continuous (particularly in $b$), there must exist a neighborhood
of $b=b_{h}$ throughout which the horizon is still surrounded by
a forbidden region.

This fact indicates that there must exist a globally disconnecting
point at $b<b_{h}$. Since we already showed that there are no ``energetic
rings'', this globally disconnecting point must be $b_{0}$ itself.
Therefore, for any $\varepsilon>0$,

\begin{equation}
b_{0}<b_{h}\,.\label{eq:b0<bh}
\end{equation}

As mentioned above, in the domain $\varepsilon>\varepsilon_{\mathrm{last}}$
(for $\alpha=0.75,\,E=\sqrt{2}$) there are two locally disconnecting
points, and we still need to find which of them corresponds to the
(single) globally disconnecting point $b_{0}$. It turns out that
only the smallest of these two $b$ values satisfies the condition
in Eq. (\ref{eq:b0<bh}) -- which signifies it as $b_{0}$.

Furthermore, note that Eq. (\ref{eq:b0<bh}) in combination with Eq.
(\ref{eq:upper bound b0}) implies $\sigma<\pi\left(b_{h}\right)^{2}$,
that is,

\begin{equation}
\sigma<\frac{4\pi Er_{+}}{\alpha\varepsilon}\equiv\sigma_{h}^{\left(+\right)}\ \ \ \ \ \ \ \ \ (\varepsilon>0)\,.\label{eq:analytical upper bound}
\end{equation}
The quantity $\sigma_{h}^{\left(+\right)}$ provides a powerful analytical
upper bound on the accretion cross section for positive $\varepsilon$
values -- see also Subsec. \ref{sec:Accretion imbalance as a function of =00003D000024=00003D00005Cvarepsilon=00003D000024}.

\section{Charge Accretion (Im)balance}
\label{chap:Results and analysis}

As explained in sections \ref{chap:Trajectory Space} and \ref{chap:Energy criterion},
the lower bound on the absorption cross section can be found by numerically
evaluating $b_{1}$, and the upper bound by finding $b_{0}$ -- or
alternatively, for repelled particles, using $b_{h}$. We implemented these methods using \textit{Wolfram
Mathematica} to compute these two bounds. Throughout this section
we focus of the specific case of $\alpha=0.75$ and $E=\sqrt{2}$.
In the first subsection we specifically consider the value $\left|\varepsilon\right|=2$
and demonstrate that charge accretion is not balanced at $Q=Q_{\mathrm{w}}$.
In the second subsection we extend the analysis to the entire range
of $\varepsilon$ and find that charge accretion imbalance persists
for all $\left|\varepsilon\right|$ values greater than a certain,
relatively small, parameter $|\varepsilon|_{\mathrm{cross}}$ to be
specified below.

The motivation for our choice of a specific $\varepsilon$ value is
simple -- as even a single example of imbalance proves that the hypothesis
of universal charge accretion balance at $Q=Q_{\mathrm{w}}$ cannot
be true. We therefore choose an explicit value ($\left|\varepsilon\right|=2$)
at which the analysis is especially simple, since it is in the domain
in which there is only a single critical point. The extension of the
analysis to the entire $\varepsilon$ axis (in Subsec. \ref{sec:Accretion imbalance as a function of =00003D000024=00003D00005Cvarepsilon=00003D000024})
is motivated by our special interest in (and the special physical
relevance of) the behavior of the charge accretion imbalance at large
$\varepsilon$ values.

For later convenience, we denote the absorption cross section for
the attracted charge as $\sigma^{-}$ and that of the repelled charge
as $\sigma^{+}$.

\subsection{A specific example of accretion imbalance}

\label{sec:Accretion imbalance at =00003D000024Q_=00003D00005Cmathrm=00003D00007Bw=00003D00007D=00003D000024}

\begin{figure}
\begin{centering}
\includegraphics[width=0.5\textwidth]{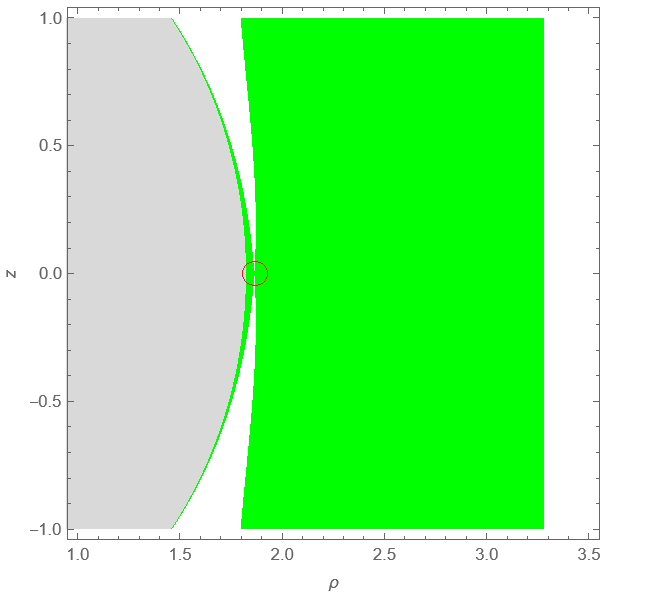} 
\par\end{centering}
\caption{Map of the allowed and forbidden regions at $b=b_{0}$ for $\alpha=0.75,E=\sqrt{2}$
and $\varepsilon=2$. For this choice of $\varepsilon$ value there is only one critical point (circled in red), which necessarily corresponds to $b_{0}$. In this case, $b_{0}\approx2.4273$. 
\label{fig:shikutz}}
\end{figure}

In this subsection we focus on the specific case $|\varepsilon|=2$
(with $E=\sqrt{2}$ and $\alpha=0.75$). 
In this case, following the method described in Subsec. \ref{sec:Critical trajectory},
we first find that for the attracted charge $b_{1}^{-}\approx4.216$.
Then, implementing the methods described in Sec.~\ref{chap:Energy criterion}
for the repelled charge, we find that at $\varepsilon=2$ there is
only a single critical point $(r_{\mathrm{crit}}\approx1.706,\theta_{\mathrm{crit}}=\frac{\pi}{2},b_{\mathrm{crit}}\approx2.427)$.
A map of the allowed and forbidden regions at $b=b_\mathrm{crit}$ is presented
in Fig.~\ref{fig:shikutz}. As explained in Subsec. \ref{sec:Globally connecting and disconnecting points} (combined with Subsec. \ref{sec:Critical points}),
in such a case, $b_{\mathrm{crit}}$ must correspond to $b_{0}$ --
namely, $b_{0}^{+}\approx2.427$. This is a relevant critical point
(namely, it satisfies $\det H<0$ and $n_{\mathrm{u},rr}<0$) which
is also locally disconnecting ($n_{\mathrm{u},b}>0$) -- as it should.

Implementing Eq.~(\ref{eq:lower bound b1}) for the attracted particle
and Eq.~(\ref{eq:upper bound b0}) for the repelled one together
with the above results for $b_{1}^{-}$ and $b_{0}^{+}$, we obtain

\begin{equation}
\sigma^{+}\leq\pi\left(b_{0}^{+}\right){}^{2}<\pi\left(b_{1}^{-}\right)^{2}\leq\sigma^{-}\,.\label{eq:accretion imbalance}
\end{equation}
The fact that $\sigma^{-}>\sigma^{+}$ (by a significant margin --
$\left(\frac{b_{1}^{-}}{b_{0}^{+}}\right)^{2}\sim3$ in this specific
case) implies that at these values of $E$, $\alpha$, and $\varepsilon$
there is a net charge accretion imbalance in favor of the attracted
charge. Therefore, $Q_{\mathrm{w}}$ cannot be the universal BH saturation
charge.
In the next subsection we will generalize the charge accretion analysis
to the entire $\varepsilon$ axis, demonstrating that the imbalance
will actually grow with increasing $\left|\varepsilon\right|$.

\subsection{Accretion imbalance as a function of $\left|\varepsilon\right|$}

\noindent \label{sec:Accretion imbalance as a function of =00003D000024=00003D00005Cvarepsilon=00003D000024}

In this subsection we generalize the analysis and results presented
in the previous subsection to the entire $\varepsilon$ axis. As already
mentioned above (see Subsec.~\ref{sec:Critical points}), to find $b_{0}$, we carry this exploration of the
$\varepsilon$ axis using two independent methods (both yielding precisely
the same results). The first one involves a dense sampling of $\varepsilon$
values along the $\varepsilon$ axis: we scanned the range $0<\left|\varepsilon\right|\leq10$
with an increment $\Delta\varepsilon=0.01$, and in addition, we scanned
the larger range $0<\left|\varepsilon\right|\leq100$ with $\Delta\varepsilon=0.1$.
The second method is more analytical in nature and requires the investigation
of only a single representative $\varepsilon$ value in each of the
domains composing the $\varepsilon$ axis (see Subsec. \ref{sec:Critical points}).

\begin{figure}
\centering \includegraphics[width=0.5\textwidth]{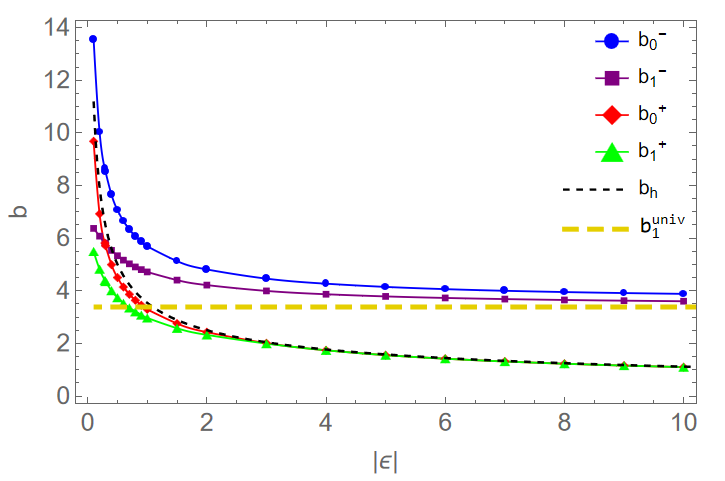}
\caption{The parameters $b_{0}^{\pm},b_{1}^{\pm}$ (which provide corresponding
bounds on the absorption cross sections $\sigma^{\pm}$) as a function
of the electromagnetic field strength parameter $\left|\varepsilon\right|$
along with the analytical curve representing $b_{h}$, and the universal
$\varepsilon\to-\infty$ limit $b_{1}^{\mathrm{univ}}$, for $E=\sqrt{2},\,\alpha=0.75$.
The curves for $b_{0}^{+}$ and $b_{1}^{-}$ intersect at $|\varepsilon|_{\mathrm{cross}}\approx0.289063$,
above which $b_{0}^{+}<b_{1}^{-}$, demonstrating accretion imbalance.
At $\varepsilon\to\infty$, $b_{0}^{+}$ and $b_{1}^{+}$ both vanish
along with $b_{h}$ while the latter serves as an analytical upper
bound on them. The curves for $b_{0}^{-}$ and $b_{1}^{-}$ converge
towards different limiting values, which, for $b_{1}^{-}$, can be
identified with $b_{1}^{\mathrm{univ}}$.}
\label{fig:main graph} 
\end{figure}

Although our imbalance analysis only requires $b_{0}^{+}$ and $b_{1}^{-}$,
we also evaluated $b_{0}^{-}$ and $b_{1}^{+}$, and all four quantities
$b_{0,1}^{\pm}$ are displayed as functions of $\varepsilon$ in Fig.~\ref{fig:main graph}.
Along with these quantities, we also display $b_{h}=\sqrt{\frac{4Er_{+}}{\alpha\varepsilon}}$
(defined for $\varepsilon>0$) and $b_{1}^{\mathrm{univ}}$.

As mentioned above, charge accretion imbalance is guaranteed if $b_{0}^{+}<b_{1}^{-}$.
From the behavior of these two quantities as presented in Fig.~\ref{fig:main graph},
we see that this inequality is indeed satisfied in the range $|\varepsilon|>|\varepsilon|_{\mathrm{cross}}\approx0.289$
-- implying that charge accretion is imbalanced in this entire range.

The domain of large $\left|\varepsilon\right|$ is of special interest
to us. In this domain, $b_{1}^{-}$ converges to the finite ($E$-independent)
value $b_{1}^{\mathrm{univ}}>0$ associated with the universal limit
$\varepsilon\to-\infty$; see Subsec. \ref{sec:Universal limit} and
in particular Fig.~\ref{fig:universal critical b1} for $b_{1}^{\mathrm{univ}}$
as a function of $\alpha$. On the other hand, as mentioned at the
end of Subsec. \ref{sec:Critical points}, $b_{0}^{+}$ decays as
$c\varepsilon^{-\frac{1}{2}}$ in the limit $\varepsilon\to+\infty$.
In our specific case, $\alpha=0.75,E=\sqrt{2}$, we find that $c\left(0.75,\sqrt{2}\right)\approx3.4727$.
This $b_{0}^{+}\propto\varepsilon^{-\frac{1}{2}}$ behavior can be
seen in Fig.~\ref{fig:main graph}. Note that for this choice of
parameters, we get $b_{h}\approx3.53996\cdot\varepsilon^{-\frac{1}{2}}$
which is (only slightly) larger than $b_{0}^{+}\approx3.4727\cdot\varepsilon^{-\frac{1}{2}}$
-- respecting the inequality in Eq. (\ref{eq:b0<bh}), as expected.
In fact, at large positive $\varepsilon$ these two curves are visually
indistinguishable in Fig.~\ref{fig:main graph}.

\begin{figure}
\centering \includegraphics[width=0.5\textwidth]{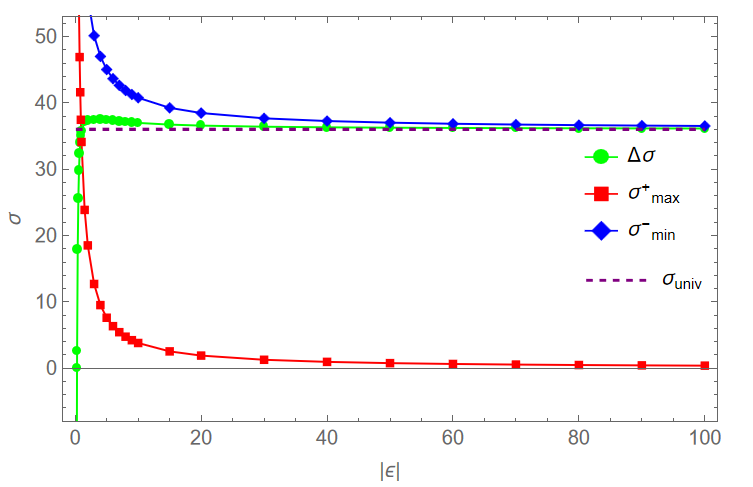}
\caption{The upper bound on the accretion cross section for repelled charges
$\sigma_{\mathrm{max}}^{+}$, the lower bound on the accretion cross
section for attracted charges $\sigma_{\mathrm{min}}^{-}$, and their
difference $\Delta\sigma=\sigma_{\mathrm{min}}^{-}-\sigma_{\mathrm{max}}^{+}$
as a function of $\left|\varepsilon\right|$, for the case $E=\sqrt{2}$,
$\alpha=0.75$. Presented as the purple dashed line is the lower bound
for attracted charges obtained through the universal EOM.}
\label{fig:imbalance graph} 
\end{figure}

We denote our upper bounds on the accretion cross sections for the
two charges as $\sigma_{\mathrm{max}}^{\pm}=\pi\left(b_{0}^{\pm}\right)^{2}$,
and the corresponding lower bounds as $\sigma_{\mathrm{min}}^{\pm}=\pi\left(b_{1}^{\pm}\right)^{2}$.
The two quantities relevant to accretion imbalance, $\sigma_{\mathrm{max}}^{+}$
and $\sigma_{\mathrm{min}}^{-}$, are displayed in Fig.~\ref{fig:imbalance graph}.
This figure demonstrates that $\sigma_{\mathrm{min}}^{-}$ goes to
a non-vanishing constant whereas $\sigma_{\mathrm{max}}^{+}$ vanishes
at the limit of large $\left|\varepsilon\right|$, as indeed follows
from the asymptotic behavior of $b_{0,1}^{\pm}$ discussed above.
As a consequence, the imbalance ratio $\frac{\sigma_{\mathrm{min}}^{-}}{\sigma_{\mathrm{max}}^{+}}$
diverges ($\propto\varepsilon$) as $\left|\varepsilon\right|$ goes
to infinity.

While out of the scope of this paper, the above analysis can be performed
for any desired pair of $\alpha,E$. In order to explore the entire
$\alpha,E$ plane we again proceeded in two independent methods (echoing
the two methods mentioned above for exploring the $\varepsilon$ axis):
(i) We explored a dense set of points covering a large domain of the
$\alpha,E$ plane, repeating the above mentioned analysis for each
pair of $\alpha$ and $E$. (ii) We developed a more systematic methodology
to identify the various \emph{phases} in the $\alpha,E$ plane. These
phases differ from each other in the number of transition points on
the $\varepsilon$ axis, and/or the properties of the critical points
in each of the $\varepsilon$ domains. We identified four such phases.
Within each phase, the properties of the sets of critical points remain
the same for all $\alpha,E$ pairs. All phases have several properties
in common: 1. The number of critical points always lies between one
and four. 2. There are no locally connecting critical points -- which
in particular implies that $b_{0}^{+}<b_{h}$ is always satisfied.
3. As already mentioned above, $b_{0}\propto\varepsilon^{-\frac{1}{2}}$
at large positive $\varepsilon$. 4. For large negative $\varepsilon$,
$b_{1}^{-}\to b_{1}^{\mathrm{univ}}\left(\alpha\right)$ (which is
displayed in Fig. \ref{fig:universal critical b1}). The last two
properties imply that charge accretion imbalance persists for all
choices of $\alpha$ and $E$, and that the imbalance ratio diverges
as $\left|\varepsilon\right|\to\infty$. We hope to further elaborate
upon that extension elsewhere.

\section{Discussion}
\label{chap:Discussion}

In this paper we have studied charge accretion into a rotating BH
embedded in an asymptotically uniform magnetic field, described by
Wald's solution \citep{Wald1974} for $A_{\mu}\left(x\right)$ (presented
in Eq. (\ref{eq:wald potential})). We set the BH charge to the value
proposed by Wald as the saturation charge for charged particle accretion,
$Q_{\mathrm{w}}=2B_{0}J$. Having in mind a dilute environment where
single-particle dynamics dominates, we compute upper and lower bounds
on charged particles' absorption cross sections and show that at least
for large enough magnetic field, the cross sections for the two types
of oppositely charged particles are not equal, implying an imbalance
in charge accretion at $Q=Q_{\mathrm{w}}$. This is our main result,
which implies that $Q_{\mathrm{w}}$ should not be thought of as the
universal model-independent saturation charge: rather, the saturation
charge depends on the specifics of the accretion model. This result
is somewhat reminiscent of the observation made in \citep{Adari2023},
although made in a different context and system (and a different setup
than the one considered by Wald), that ``dynamics upstage screening''.
Nevertheless, at least in an important (and likely astrophysically
relevant) region of the parameter space of the model, Wald's charge
$Q_{\mathrm{w}}$ still seems to provide a good approximation to the
actual saturation charge, as we further discuss below.

For both positive and negative particle charge, there is a ``central
absorption disk'' with cross sectional radius $b_{1}$ such that
all orbits with $b<b_{1}$ fall into the BH. We solved the EOM numerically
for both attracted and repelled charges and found $b_{1}$ for various
choices of $\varepsilon$ (as well as $E$ and $\alpha$) providing
the \emph{lower} \emph{bound} $\pi\left(b_{1}\right){}^{2}$ on the
absorption cross section. We found that in the case of an attracted
charge, $b_{1}^{-}$ tends toward a constant value of order a few
times $M$ at large dimensionless field strength $\left|\varepsilon\right|$,
depending only on the spin parameter $\alpha$. This was found by
using the $\varepsilon\to-\infty$ universal limit of the EOM introduced
in Subsec.~\ref{sec:Universal limit}, where the trajectories become
formally light-like and independent of $E$ (and, obviously, of $\varepsilon$). This limit (which was also numerically verified; see Fig.~\ref{fig:convergence to universal b1})
provides enhanced -- yet incomplete -- analytical control which
improves our global understanding of the critical trajectory $b_{1}^{-}$.
In contrast, for the repelled charge we find that for $\varepsilon\gg1$
the corresponding critical trajectory converges towards the axis of
symmetry $\rho=0$, and in particular, the impact parameter $b_{1}^{+}$
decreases to zero.

On the other hand, the CQB criterion introduced in Sec.~\ref{chap:Energy criterion}
provides a method for obtaining \emph{upper} \emph{bounds} on the
absorption cross section for charges of both signs. We use this criterion
to identify spatial regions which are disallowed for the presence
of the charged particle as a function of $b$. A particle with a given
$b$ is energetically allowed to be absorbed if and only if there
is an allowed spatial region connecting the neighborhood of $\rho=b$
at large $z$ to the horizon. By analyzing the critical points on
the $b$ axis, in which the allowed region changes its topology, we
identify $b_{0}$ -- namely, the maximal impact parameter value for
which absorption is energetically allowed -- which provides an upper
bound $\pi\left(b_{0}\right){}^{2}$ on the absorption cross section.
More generally, we find that the analysis of critical points (in the
space of $b,r$ and $\theta$) provides insight into the space of
possible orbits for fixed system parameters. In terms of their number
and characteristics, the critical points display a rich structure
which differs qualitatively in different regions of the system's parameter
space. This structure is partially discussed here, and we hope to
further elaborate on it in a following paper. Finally, we show that
the CQB criterion analysis significantly simplifies at $\varepsilon\gg1$,
and in particular that $b_{0}^{+}\propto\varepsilon^{-\frac{1}{2}}$
in this regime, providing a vanishing upper bound on the cross section
for repelled charges.

The accretion situation for the case of large $\varepsilon$ therefore
becomes clear: The absorption cross section of the attracted particle
is bounded below by a few times $M^{2}$, whereas that of the repelled
particle vanishes as fast as $\varepsilon^{-1}$. This unavoidably
leads to a non-vanishing net charge accretion at $Q=Q_{\mathrm{w}}$.
In fact, this accretion imbalance is not limited to the large-$\varepsilon$
limit -- it already occurs at $\varepsilon$ values of order unity,
as can be seen in Figs.~\ref{fig:main graph},\ref{fig:imbalance graph}.

\begin{figure}
\centering \includegraphics[width=0.4\textwidth]{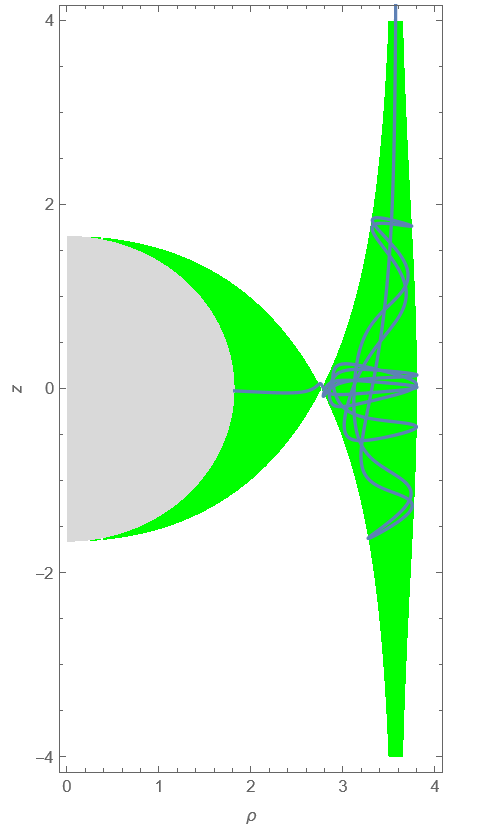}
\caption{The trajectory of an attracted charged particle (blue). The particle
is confined to the allowed region (green), bouncing back and forth
until it eventually falls into the BH. The trajectory presented here
has parameters $E=\sqrt{2},\alpha=0.75,\varepsilon=-1000$ and $b=3.5797444$.
It has a $b$ value very close to its corresponding $b_{0}$.}
\label{fig:sauron} 
\end{figure}

As was expanded upon in Sec.~\ref{chap:Trajectory Space}, for $b<b_{1}$
all particles fall into the BH, while for $b>b_{0}$ particles are
energetically incapable of falling into the BH. The region between
these bounds, $b_{1}<b<b_{0}$, generally shows complex behavior,
and the domain on the $b$ axis corresponding to absorption seems
to be of fractal nature. This fractal behavior follows from the interplay
between several basic characteristics of the system, as we hope to
show elsewhere. This fractal structure is also reflected in the complex
behavior of typical trajectories within this domain of $b$: these
orbits generically bounce many times before either falling into the
BH or escaping to infinity (or perhaps remaining trapped indefinitely)
as demonstrated in Fig.~\ref{fig:sauron}. A fractal behavior of
similar nature has been described in the literature in somewhat different
contexts (e.g. considering different classes of orbits)\citep{Frolov2013,Adari2023,Levin2018}.

Upon close inspection of the behavior of trajectories at $\varepsilon\rightarrow-\infty$
(using analytical as well as numerical tools), we hypothesize that
in this limit all attracted particles with $b<b_{0}^{-}$ will eventually
fall into the BH (except a measure-zero set), even if they bounce
around in the allowed region for a long time. As a consequence, in
this limit, the energetic upper bound $b_{0}^{-}$ provides the true
absorption cross section for the attracted particle: $\pi\left(b_{0}^{-}\right){}^{2}$.
We hope to expand upon this subject in the future. 
Since the parameter $\varepsilon$ also depends on the charge-to-mass
ratio of the particles being accreted, it is important to mention
that while we assumed the same charge-to-mass ratio for attracted
and repelled particles, the analysis straightforwardly generalizes
to the case of different charge-to-mass ratios, and ultimately the
result of accretion imbalance at $Q_{\mathrm{w}}$ is general. Namely,
in a more plausible scenario involving ionized hydrogen -- electrons
and protons -- the difference in mass will result in one of the curves
(either $\sigma_{\mathrm{max}}^{+}$ or $\sigma_{\mathrm{min}}^{-}$)
presented in Fig.~\ref{fig:imbalance graph} being horizontally stretched
relative to the other curve (along the $|\varepsilon|$ axis). For
sufficiently large values of the asymptotic magnetic field $B_{0}$,
$\sigma_{\mathrm{max}}^{+}$ still decreases to zero proportionally
to $B_{0}^{-1}$, whereas $\sigma_{\mathrm{min}}^{-}$ approaches
a constant of order a few times $M^{2}$ as discussed above. Therefore,
accretion imbalance still occurs in this more realistic situation\footnote{When considering particles of different masses, the assumption of
equal energy per unit mass $E$ is not necessarily the most natural
(e.g. when the electrons and the ions share the same temperature).
Nevertheless, the vanishing of $\frac{\sigma_{\mathrm{max}}^{+}}{\sigma_{\mathrm{min}}^{-}}$
at the large-$\left|\varepsilon\right|$ limit is still guaranteed,
implying accretion imbalance at $Q=Q_{\mathrm{w}}$. This argument
also holds for two constituents of different charge magnitudes (like
fully ionized Helium).}.

Since at the limit of large $\left|\varepsilon\right|$ the imbalance
ratio diverges, one might naively expect that at this limit the actual
saturation charge $Q_{\mathrm{sat}}$ should significantly differ
from $Q_{\mathrm{w}}$. However, a closer inspection reveals this
not to be the case. In a forthcoming paper \citep{Okun2025}, we will
consider the correction to $Q_{\mathrm{w}}$ at large $|\varepsilon|$
within our accretion model, and show that the relative saturation
charge correction $\delta\equiv(Q_{\mathrm{sat}}-Q_{\mathrm{w}})/Q_{\mathrm{w}}$
actually takes the form $\delta=\zeta|\varepsilon|^{-\frac{2}{3}}$
where $\zeta(\alpha,E)$ is an $\varepsilon$-independent prefactor.
Thus, $Q_{\mathrm{w}}$ remains the leading-order saturation charge
when $|\varepsilon|\gg1$. To illustrate the astrophysical relevance
of this regime, it is instructive to estimate $|\varepsilon|$ for
electrons being accreted by Sgr A{*}. For this case, we can write
$|\varepsilon|=\left|\frac{e}{m}B_{0}M\right|\approx\frac{|e|}{m}\cdot0.85\cdot10^{-20}\frac{M}{M_{\odot}}\frac{B_{0}}{\mathrm{Gauss}}$.
Plugging in the estimated BH mass $4\cdot10^{6}M_{\odot}$ \citep{Gravitycollab2023}
and its surrounding magnetic field $\sim10^{2}\ \mathrm{Gauss}$ \citep{Gravitycollaboration2020}
along with the charge-to-mass ratio of the electron, we obtain $\varepsilon\sim3\cdot10^{9}$
(and $\varepsilon\sim10^{6}$ for protons).

These estimates demonstrate that the limit $|\varepsilon|\gg1$ considered
in our analysis lies well within the range expected near astrophysical
BHs, motivating its physical relevance. More generally, this study
offers insights into some basic mechanisms that may govern charge
accumulation in rotating BHs. Such understanding may have important
implications for high-energy astrophysical phenomena near BHs.

\acknowledgements
We are grateful to Noa Zilberman for helpful comments. A. Ori would like to thank Bob Wald for helpful discussions. This work was supported in part by the Israel Science Foundation (grant No. 2047/23).


\appendix

\section{Phases of the Critical Trajectory $b_1$} 
\label{app:critical trajectory}

\begin{figure}
\centering \includegraphics[width=0.5\textwidth]{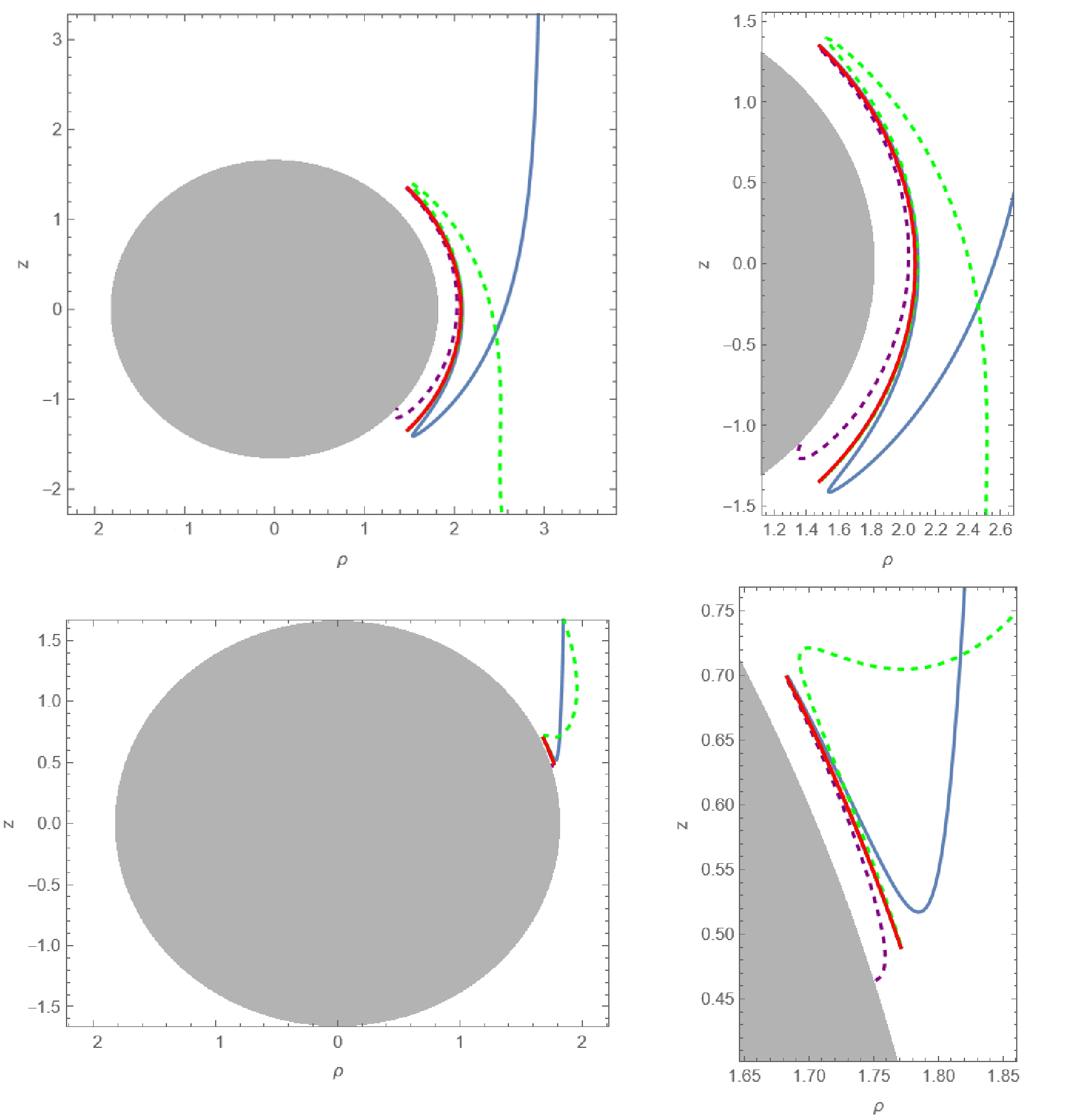}
\caption{The two different phases of the critical trajectory. The second column
is a closeup of the first, focused on the neighborhood of the asymptotic
oscillatory trajectory, further explained below. The first row represents
the phase $\varepsilon<\varepsilon_{\mathrm{crit}}$, where the critical
trajectory oscillates, asymptotically symmetrically, around the equatorial plane. The
second row represents the phase $\varepsilon>\varepsilon_{\mathrm{crit}}$,
where the critical trajectory oscillates without crossing the equatorial
plane. The solid blue line represents the incoming critical trajectory $b=b_1$, which at late times asymptotically approaches a precisely periodic trajectory (the solid red curve).
The two dashed lines show near-critical orbits coming from infinity with
$b\gtrless b_{1}$ along trajectories very close to the blue one, which either escape to infinity (green) or
fall into the BH (purple) after oscillating several times near the red curve.} 
 \label{fig:critical trajectory oscillations} 
\end{figure}

\begin{figure}
\centering \includegraphics[width=0.5\textwidth]{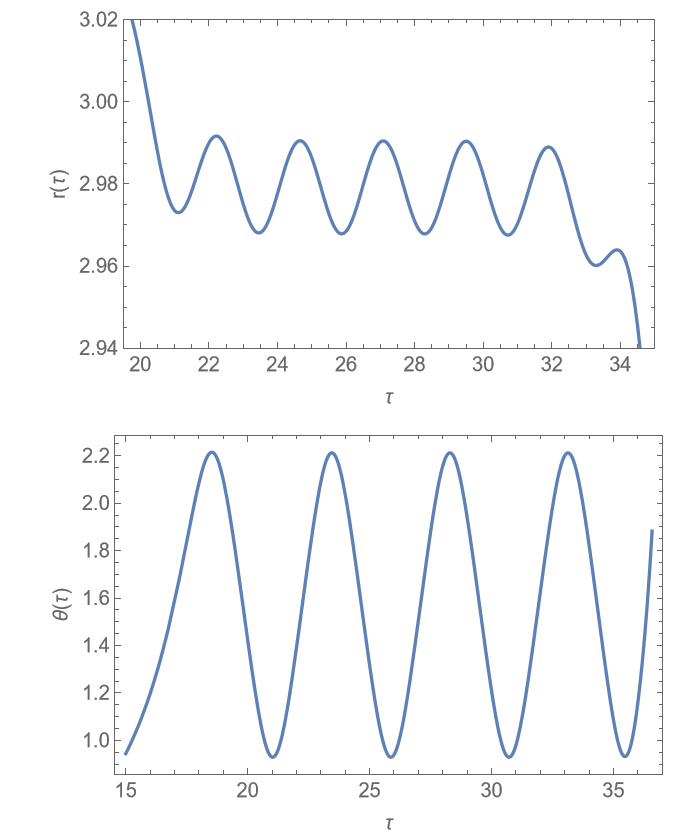}
\caption{Oscillations of $r$ (top panel) and $\theta$ (bottom panel) as a
function of the proper time $\tau$. The trajectory presented here
is an infalling trajectory which is very close to the critical one,
with $E=\sqrt{2},\alpha=0.5,\varepsilon=-1$,  and $b=4.450818236305003$. }

\label{fig:r and theta oscillations} 
\end{figure}

While the critical trajectory $b=b_1$ is oscillatory for all choices of $E$
and $a$, we observe that there is a critical value of $\varepsilon$
which separates the space of critical trajectories into two qualitatively
different phases, depending only on the value of $\varepsilon$. These
two phases are characterized by an oscillation that is symmetric with
respect to the equatorial plane for $\varepsilon<\varepsilon_{\mathrm{crit}}$,
and an oscillation that occurs only above the equatorial plane in
the case $\varepsilon>\varepsilon_{\mathrm{crit}}$, as can be seen
in Fig.~\ref{fig:critical trajectory oscillations}. It is important
to note that in both phases, both the $r$ and $\theta$ coordinates
oscillate, as can be seen in Fig.~\ref{fig:r and theta oscillations}.
In the limit of $\varepsilon\rightarrow\varepsilon_{\mathrm{crit}}$,
the critical trajectory is expected to converge towards a circular
orbit on the equatorial plane.

When $\varepsilon\rightarrow\infty$, the amplitude of the oscillations
in both $r$ and $\theta$ tends toward zero and $b_{1}$ approaches
zero. This is because the strength of the electric potential is proportional
to $\varepsilon$ and as such, any charge coming towards the BH will
be repelled well before it comes close enough to be affected by the
gravitational field, except on the axis of symmetry where the electric
potential vanishes.


\section{Demonstration of $n_{,r}|_{r=r_{+},b=b_{h}}>0$} \label{app:nrposdem}

\label{Demonstration of the positivity of $n_{,r}|_{r=r_{+},b=b_{h}}$}
In this appendix we demonstrate the fact stated in Subsec.~\ref{sec:An analytical upper bound}, regarding the positivity of $n_{\mathrm{u},r}$ when evaluated at $r=r_+,b=b_h$. This implies the existence of a forbidden region surrounding the BH, which is central to justifying the use of $\pi (b_h)^2$ as an upper bound on the accretion cross section for repelled particles.

Here we will restrict ourselves to subextremal BHs, $\alpha<1$, and moreover
assume below that $\alpha>0$ since $b_{h}$ diverges at $\alpha=0$.
We denote $v\equiv\sin^{2}\theta$. 

We begin by evaluating $n_{,r}$ at $r=r_{+}$ and $b=b_{h}$ \eqref{eq:bh}:
\begin{align}
n_{\mathrm{u},r}|_{r=r_{+},b=b_{h}}=\frac{32E^{2}}{\alpha^{2}}r_{+}^{2}\left(r_{+}-1\right)+\frac{16v}{\alpha}r_{+}\left(r_{+}-1\right)\nonumber \\
\times\left(-2E^{2}\alpha-2E\varepsilon r_{+}+\alpha\right)+8v^{2}\left(r_{+}-1\right)\nonumber \\
\times\left(2E\varepsilon\alpha r_{+}+\left(E^{2}-1\right)\alpha^{2}+\varepsilon^{2}r_{+}^{2}\right)\,.\nonumber \\
\end{align}
Our goal is to show that $n_{\mathrm{u},r}>0$ for $0\leq v\leq1$.
We first consider the boundaries of this range. At $v=0$, we find
\begin{align}
n_{\mathrm{u},r}|_{r=r_{+},b=b_{h},v=0}=\frac{32E^{2}}{\alpha^{2}}r_{+}^{2}\left(r_{+}-1\right)>0\,.
\end{align}

At $v=1$, we find 
\begin{align}
n_{\mathrm{u},r}|_{r=r_{+},b=b_{h},v=1}=\frac{8}{\alpha^{2}}r_{+}^{2}\left(r_{+}-1\right)\left(\alpha^{2}+\left(Er_{+}-\alpha\varepsilon\right)^{2}\right)~.\label{eq:-1}
\end{align}
This expression is clearly positive, and cannot vanish for $0<\alpha<1$.
Therefore, 
\begin{align}
n_{\mathrm{u},r}|_{r=r_{+},b=b_{h},v=1}>0.
\end{align}
We have thus shown that $n_{\mathrm{u},r}$ is positive at the boundaries
$v=0,1$. Now, if $n_{\mathrm{u},r}$ is (weakly) monotonic in the
range $0\leq v\leq1$, the positivity of $n_{\mathrm{u},r}$ follows
immediately. Otherwise, since $n_{\mathrm{u},r}$ is a quadratic polynomial
in $v$, it suffices to show that at its extremum, which occurs
at some $0<v_{\mathrm{ext}}<1$, it assumes a positive value. 
Solving $\left(n_{\mathrm{u},r}\right){}_{,v}|_{v_{\mathrm{ext}}}=0$
yields 
\begin{align}
v_{\mathrm{ext}}=\frac{\alpha\left(2E^{2}-1\right)r_{+}+2Er_{+}^{2}\varepsilon}{\alpha\left(\left(\alpha E+\varepsilon r_{+}\right)^{2}-\alpha^{2}\right)}\,,
\end{align}
and plugging back into $n_{\mathrm{u},r}$ gives 
\begin{align}
n_{\mathrm{u},r}^{\mathrm{ext}}=\frac{8r_{+}^{2}\left(r_{+}-1\right)\left(4Er_{+}\varepsilon-\alpha\right)}{\alpha\left(\left(\alpha E+\varepsilon r_{+}\right)^{2}-\alpha^{2}\right)}\,,\label{eq:n,r_ext}
\end{align}
whose denominator is evidently always positive.

To complete our proof of the positivity of $n_{\mathrm{u},r}$, we
must show that $n_{\mathrm{u},r}^{\mathrm{ext}}$ is always positive
when $0<v_{\mathrm{ext}}<1$. By evaluating $n_{\mathrm{u},r}^{\mathrm{ext}}$
at, for example, $\varepsilon=50$, $E=2$ and $\alpha=0.1$, we show
that there exists at least one point in the space parameterized by
$\varepsilon$, $E$, and $\alpha$, for which $n_{\mathrm{u},r}^{\mathrm{ext}}$
is positive and $0<v_{\mathrm{ext}}<1$. Let's suppose that there
exists another point for which $n_{\mathrm{u},r}^{\mathrm{ext}}<0$
and $0<v_{\mathrm{ext}}<1$. Since we showed that at the boundaries
$v=0,1$, $n_{\mathrm{u},r}>0$, and that the denominator of $n_{\mathrm{u},r}^{\mathrm{ext}}$
is positive, by continuity there must exist values of $\varepsilon$,
$E$, and $\alpha$, for which $n_{\mathrm{u},r}^{\mathrm{ext}}=0$
and $0<v_{ext}<1$. Let us find the value of $\varepsilon=\varepsilon_{0}$
for which $n_{\mathrm{u},r}^{\mathrm{ext}}=0$ for any $E$ and $\alpha$.
From \eqref{eq:n,r_ext}, we get 
\begin{align}
\varepsilon_{0}=\frac{\alpha}{4Er_{+}}\,.
\end{align}
Evaluating $v_{\mathrm{ext}}$ at $\varepsilon_{0}$ gives 
\begin{align}
v_{\mathrm{ext}}^{0}=\frac{8E^{2}r_{+}}{\alpha^{2}\left(4E^{2}-1\right)}\,.
\end{align}
Assuming $v_{ext}^{0}<1$, then, implies 
\begin{align}
E^{2}<-\frac{\alpha^{2}}{4r_{+}^{2}}\,,
\end{align}
and therefore no real $E$ exists for which $v_{\mathrm{ext}}^{0}<1$.
This shows that $n_{\mathrm{u},r}^{\mathrm{ext}}$ cannot vanish when
$0<v_{\mathrm{ext}}<1$ and, as such, $n_{\mathrm{u},r}^{\mathrm{ext}}>0$
in that case. 
We conclude that 
\begin{align}
n_{\mathrm{u},r}|_{r=r_{+},b=b_{h}}>0\,,
\end{align}
for all $\alpha$, $E$, $\varepsilon$, and $0\leq v\leq1$.

\section{The large $\varepsilon$ limit of $b_0^{-}$}
\label{b0m limit}

Although not directly relevant for the main claim
of this paper, here we present another application of the universal
limit $\varepsilon\to-\infty$ (see Subsec.~\ref{sec:Universal limit}) --
namely, the (semi-)analytical computation of the upper bound
$\pi\left(b_{0}^{\mathrm{univ}}\right)^{2}$ on the large-$\left|\varepsilon\right|$
absorption cross section of an attracted particle, where
$b_{0}^{\mathrm{univ}}\equiv\lim_{\varepsilon\to-\infty}b_{0}^{-}(\varepsilon)$.
The CQB procedure implemented (in
Sec.~\ref{chap:Conserved Quantities Based (CQB) Criterion}) at finite $\varepsilon$ for finding the critical points
of the normalization function $n$ (Eq.~\eqref{eq:cqb criterion}) is directly applicable
to the normalization condition (Eq.~\eqref{eq:universal normalization-2}) associated with the universal $\varepsilon\to-\infty$
limit. It yields a set of equations analogous to
Eqs.~\eqref{eq:critical equations 1},~\eqref{eq:critical equations 2},~\eqref{eq:critical equations 3}, allowing for a direct computation of
the limiting value $b_{0}^{\mathrm{univ}}$. The finite-$\varepsilon$ values
calculated for $b_{0}^{-}$ (using the methods described
in Sec.~\ref{chap:Conserved Quantities Based (CQB) Criterion}) show excellent convergence to this limiting value --
in analogy to what was shown for $b_{1}^{-}$ in Fig.~\ref{fig:convergence to universal b1}.

However, we found that the value of $b_{0}^{\mathrm{univ}}$
may be obtained through a computationally simpler procedure, as we
now sketch.

The normalization in the universal limit attains a remarkably simple
form: 
\begin{align}
g^{\phi\phi}b^{4}-2b^{2}+g_{\phi\phi}=0~.\label{eq:universal normalization}
\end{align}

When considering a particle as it crosses the equatorial plane, Eq.~(\ref{eq:universal normalization}) becomes even simpler. For any
given value of $r>2$, this polynomial in $b$ has
two real positive roots which we denote $b_{\mathrm{root},i}(r)$ (where $i=1,2$). These roots then
dictate the range of $b$ values that would allow the particle to arrive at the
equatorial plane at that $r$ value. Upon
further investigation one finds that the larger root,
$b_{\mathrm{root},2}(r)$, attains a minimum value at
some $r\equiv r^{\mathrm{min}}>2$ -- which we denote $b_{\mathrm{root},2}^{\mathrm{min}}$.
Moreover, we find \cite{Okunthesis}
that this $b$ value actually corresponds to the maximal impact parameter
for which the orbit is energetically allowed to
cross the surface defined by $r=2$ (at any $\theta$). Consequently,
every trajectory with $b>b_{\mathrm{root},2}^{\mathrm{min}}$ is obviously
energetically disallowed to fall into the BH (since $r_{+}\leq2$
for all $\alpha$). As it turns out, $b_{\mathrm{root},2}^{\mathrm{min}}$ proves to be equal to $b_{0}^{\mathrm{univ}}$\footnote{This equality reflects the fact that in the context
of the universal limit, the critical point associated with $b_{0}^{\mathrm{univ}}$
is equatorial.}. This value, which depends only on $\alpha$, may
be presented compactly by:  
\begin{equation}
\left(b_{0}^{\mathrm{univ}}\right)^{2}=\min_{r}\frac{r\Delta+2\alpha\sqrt{\Delta}}{r-2}~,\label{eq:  b_0^univ}
\end{equation}
where, recall, $\Delta\equiv r^{2}-2r+\alpha^{2}$. Equation (\ref{eq:  b_0^univ})
allows us to directly obtain $b_{0}^{\mathrm{univ}}$ in a simple (semi-)analytical
manner.

\section{An example of the classification process of critical points}

\label{app:classification}

We here demonstrate the implementation of our critical point classification
process -- and thereby the identification of $b_{0}$ -- in the
last $\varepsilon$ domain (namely, $\varepsilon>\varepsilon_{\mathrm{last}}\approx6.8702$)
for $\alpha=0.75,E=\sqrt{2}$, as described in subsections \ref{sec:Critical points}
and \ref{sec:An analytical upper bound}.  

As discussed in Sec.~\ref{chap:Conserved Quantities Based (CQB) Criterion}, the critical points can be divided into two categories: equatorial points and non-equatorial points. For a critical point to be considered as relevant, the sign of the determinant of the Hessian matrix of $n_\mathrm{u}$ evaluated at this point must be negative. Furthermore, for equatorial points, the additional condition $n_{\mathrm{u},rr}<0$ must be met. After the relevant points have been identified, we seek out the points which are locally disconnecting, i.e. $n_{\mathrm{u},b}>0$. Finally, to discern between remaining candidates, we identify as corresponding to $b_0$ the critical point whose $b$ value is smaller than $b_h$.

In this domain there are only four critical points. One of them is non-equatorial, while the other three are on the equatorial plane. While their actual $r,\theta$ and $b$ values change as a function of $\varepsilon$, their ordering and the properties relevant to the classification process remain unchanged throughout the entire domain. We find that for these values of $\alpha$ and $E$, the critical point corresponding to $b_0$ is always the single non-equatorial one. A summary of the classification process is presented in Table~\ref{table}.

\begin{table}
\begin{centering}
\includegraphics[width=0.5\textwidth]{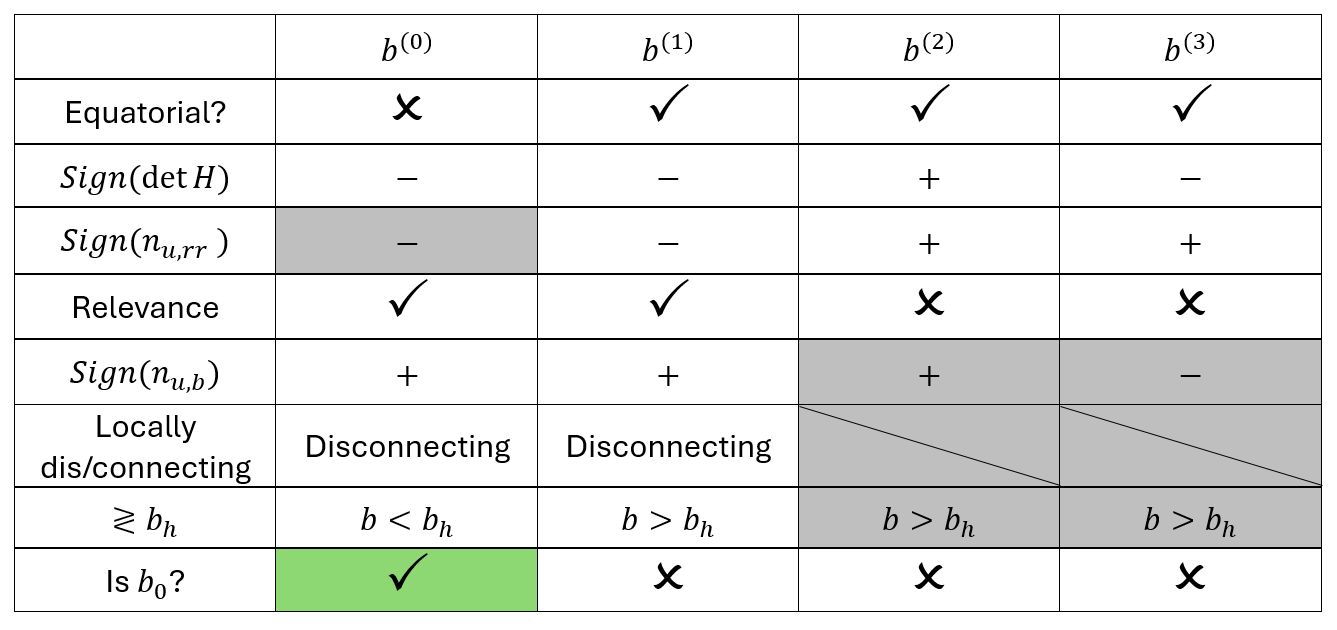}
\par\end{centering}
\caption{Classification of critical points in the domain of $\varepsilon>\varepsilon_{\mathrm{last}}\approx6.8702$
for $\alpha=0.75$ and $E=\sqrt{2}$. The critical points are represented by their $b$ values and in ascending order with respect to them. 
Each row represents a property of the critical points.
The cells in gray are properties that, while remaining unchanged throughout the $\varepsilon$-domain, are irrelevant to the identification of $b_0$. The last two points may be disqualified (for $b_0$ correspondence) early on as they are not relevant points. The first point is non-equatorial and therefore, the sign of $n_{\mathrm{u},rr}$ is irrelevant to its classification. The point which corresponds to $b_0$ is highlighted in green.}

\label{table}
\end{table}

\bibliography{references.bib}

\end{document}